\documentclass[a4paper,11pt]{article}
\pdfoutput=1
\usepackage{jinstpub}
\usepackage{graphicx}
\usepackage{subcaption}
\usepackage{newtxmath}
\usepackage{ulem}

\usepackage{array}
\usepackage{lineno}
\linenumbers
\newcommand{\PreserveBackslash}[1]{\let\temp=\\#1\let\\=\temp}
\newcolumntype{C}[1]{>{\PreserveBackslash\centering}p{#1}}

\author[a]{H.~Pernegger \footnote{Corresponding author, \tt{Email: heinz.pernegger@cern.ch}}}
\author[a]{E.~K.~Anderson}
\author[b]{P.~Bartulovi\' c}
\author[b]{I.~Berdalovi\' c}
\author[a]{M.~Giroux de Foiard Brown}
\author[a,c]{S.~Haberl}
\author[b]{M.~Jugovi\' c}
\author[a]{A.~Kotsokechagia}
\author[a,d]{J.~Lunde}
\author[b]{B.~Po\v zar}
\author[b]{T.~Suligoj}

\affiliation[a]{CERN, Experimental Physics Department, Geneva, Switzerland}
\affiliation[b]{University of Zagreb, Faculty of Electrical Engineering and Computing, Zagreb, Croatia}
\affiliation[c]{University of Innsbruck, Faculty of Engineering Science, Innsbruck, Austria}
\affiliation[d]{University of Oslo, Department of Physics, Oslo, Norway}

\title{First results of a Monolithic Active Pixel Sensor with Internal Signal Gain Fully Integrated in a 180~nm CMOS Technology}

\abstract{Dense tracking environments in experiments at CERN's High-Luminosity LHC and future FCC experiments call for an increased use of timing information in addition to the position measurement of pixel detectors. This adds one dimension to the information available, and is essential for pile-up mitigation at high luminosity. The CASSIA sensor project ({\bf C}MOS {\bf A}ctive {\bf S}en{\bf S}or with {\bf I}nternal {\bf A}mplification) focuses on the development of pixel matrices with internal charge multiplication, based on monolithic CMOS sensor technologies, suitable for application as charged particle tracking and timing detectors.  CMOS sensors with in-pixel internal amplification would provide higher signal amplitudes, an improved signal-to-noise ratio, better time resolution and increased sensitivity, making them attractive for high-radiation environments. Their monolithic integration in small pixels reduces the parallel capacitance presented to a front-end amplifier and the power dissipation making it suitable for fine-pitch low-power detectors. Fast signal rise time due to internal charge amplification improves the response time and timing resolution, all of which make such a technology attractive for future 4D tracking applications in HEP experiments. This paper presents the first results of the CASSIA sensor, a novel MAPS which uses gain layers fully integrated in a 180nm imaging process to achieve internal signal amplification. In the first measurements presented here we demonstrate the gain behaviour of different pixel implant designs and show that the sensor can be operated at low gain proportional mode as LGAD sensor at lower voltages and as SPAD sensor at higher voltages .}


\date{\today}
\nolinenumbers
\notoc

\begin{document}

\maketitle

\section{Introduction}
Monolithic active pixel sensors (MAPS) for charged particle detection offer a compact structure that reduces material budget and decreases parasitic capacitances at the sensor output node, resulting in lower noise and power dissipation. They are also more cost-effective than hybrid detectors for large tracking systems, eliminating the need for separate sensor and readout chips as well as expensive bump-bonding interconnects. Most monolithic pixel detectors demonstrated to date utilize passive, predominantly pin-diode based sensors, where the charge signal generated by incident particles is amplified by an integrated front-end amplifier.

In contrast, low-gain avalanche diodes (LGADs)~\cite{FirstLGAD} amplify the primary signal through avalanche multiplication within a p-n junction before electronic amplification, providing higher signal amplitudes for a given number of generated charge carriers. LGADs are currently implemented as chips separated from the signal-processing electronics, using specialized fabrication technologies optimized to obtain a uniformly distributed avalanche multiplication region. This design captures charge carriers generated at any position within the entire pixel area, providing a high fill factor~\cite{LaterLGAD}. LGAD devices are typically fabricated separately because they require specialized implant structures and high-resistivity substrates that differ from standard CMOS processes used for readout electronics.
The CMOS Active Sensor with Internal Amplification (CASSIA) project addresses this technological incompatibility by exploring novel designs that integrate avalanche multiplication directly into a high-volume monolithic CMOS imaging process. This approach combines the benefits of monolithic pixel sensors with internal signal amplification in a single device. By amplifying the primary ionization signal inside the pixel, CASSIA sensors provide a larger current signal, which significantly improves performance compared to current state-of-the-art radiation-hard fine-pitch CMOS pixel sensors for high-energy physics (HEP) applications. The key advantages include:
\begin{itemize}
\item Higher signal amplitudes that enable simplification of in-pixel electronics, contributing to lower power consumption,

\item Enhanced timing resolution in fine-pitch MAPS for future 4D tracking and time-tagging applications,

\item Improved signal-to-noise ratio for operation in high-radiation environments.
\end{itemize}

The CASSIA sensor is designed to provide two distinct operation modes with internal charge multiplication. In low-gain avalanche diode (LGAD) mode, the sensor operates below the hard breakdown voltage with moderate gain (typically 10--100), providing low noise rates essential for many applications. Operation at or above the hard breakdown voltage enables Geiger mode operation as a single-photon avalanche diode (SPAD), offering the highest time resolution~\cite{spad_timing}. SPADs provide extremely high intrinsic gain that can eliminate the need for a front-end amplifier, further improving the signal-to-noise ratio. However, SPAD operation presents challenges: devices require quenching and reset circuits to operate above breakdown voltage~\cite{spad_zappa}. Additionally, SPADs investigated for particle detection applications ~\cite{spad_particle} exhibit limitations including higher dark count rates (DCR) when integrated in CMOS~\cite{spad_cmos} and DCR degradation in high-radiation environments~\cite{Wu2023}.

In the full CMOS integration pursued for CASSIA, the same process can fabricate monolithic sensors without gain, LGADs, and SPADs, sharing common circuit blocks for signal processing, pixel address generation, and matrix readout. The operation mode is selected by adjusting the reverse bias voltage applied to the sensor, which controls the avalanche multiplication rate.

The implementation of LGAD and SPAD sensors in a commercial CMOS imaging is of great interest to particle physics for application as timing or tracking detector as well as for applications in low-energy X-ray detection and for astrophysics instrumentation. While these applications share strong interest for state-of-the-art imaging devices with internal gain, their specific requirements impact the sensor's design choices. In the CASSIA project we explore different designs and processing choices through variation in pixel fillfactor, choice of substrate and active volume thickness, gain layer doping and implant geometries which enable LGAD- and SPAD-mode operation. 

The results of this development will allow future sensor designs to be optimized towards their intended application. Tracking detectors typically require 100\% single-particle detection efficiency and very low noise rates. They therefore favour designs with large fillfactor operating in LGAD mode, where noise signals can be suppressed by setting an appropriate charge threshold. Timing detectors may not require 100\% fillfactor but benefit from thin active volumes and high operational voltages in SPAD mode for optimal timing performance. However, they may be able to tolerate higher dark count rates in SPAD mode, where thermally generated electron-hole pairs produce noise signals that are indistinguishable from those originating from incident photons or charged particles.

The CASSIA project also explores different gain layer doping profiles in view of applications requiring tolerance to radiation damage from non-ionizing energy loss (NIEL), which effects the gain through changes in p-type gain layer doping. 

The CASSIA1 prototype has been manufactured on 30$\,\upmu\text{m}$ thick epitaxial substrates as well as thick high-resistivity Czochralski substrates with an expected depleted thickness approaching $100\,\upmu\text{m}$. A thick depleted volume is beneficial for low-energy X-ray detection and for operation in high-radiation environments, while a thin active volume is beneficial for optimal timing performance. 

This paper focuses on results obtained on the CASSIA1 prototype sensor: expected and measured bias voltages for LGAD and SPAD amplification as function of electrode and gain layer doping configuration as well as function of gain layer size. Furthermore we analyse signal characteristics in response to laser pulses and compare dark count-rates for different designs. Initial studies on time-resolution, performance changes under irradiation and substrate comparisons are currently ongoing on CASSIA1 prototypes and are subject to future publication. Optimization for maximum efficiency close to 100\% as well as extended radiation hardness studies will be addressed in future developments using the CASSIA2 sensor, which is currently being manufactured and includes different versions of in-pixel front-end electronics which are optimized for LGAD and for SPAD operation.

In this work, we explore the implementation of internal amplification in the depleted monolithic pixel sensor technology used for the MALTA sensor family, which has demonstrated state-of-the-art performance in particle detection applications~\cite{Pernegger_2023, dyndal2019}. MALTA sensors exhibit particle detection efficiencies of 98--99\% after exposure to non-ionizing energy loss (NIEL) fluences of $3\times10^{15}$~1~MeV~n$_{\text{eq}}$/cm$^2$, with timing resolution at the nanosecond level. Critically, this integration requires no changes to process parameters and uses only layers already available in the standard fabrication sequence. This research aims to combine typical HEP pixel detector pitches (50--100$\,\rm \upmu$m) with timing resolution sufficient to enable future 4D tracking applications (target:~$<$100\,ps)~\cite{4dLGAD}.

To maintain radiation hardness at comparable fluences and further improve timing performance, the primary goal of the CASSIA project is full integration into the existing 180\,nm TowerSemiconductor imaging sensor technology used for MALTA fabrication. In particular, changes to p-type gain layer doping with NIEL irradiation will be investigated in the future on CASSIA sensors. The current CASSIA1 prototype uses higher p-type gain layer doping concentration than typical LGAD detectors. This can mitigate gain decrease under radiation and extend sensor's radiation hardness but may impact gain layer depletion for optimal charge collection, related studies are currently on-going.


\section{The CASSIA sensor design} \label{design}

\begin{figure}
    \centering
    \includegraphics[width=0.95\textwidth]{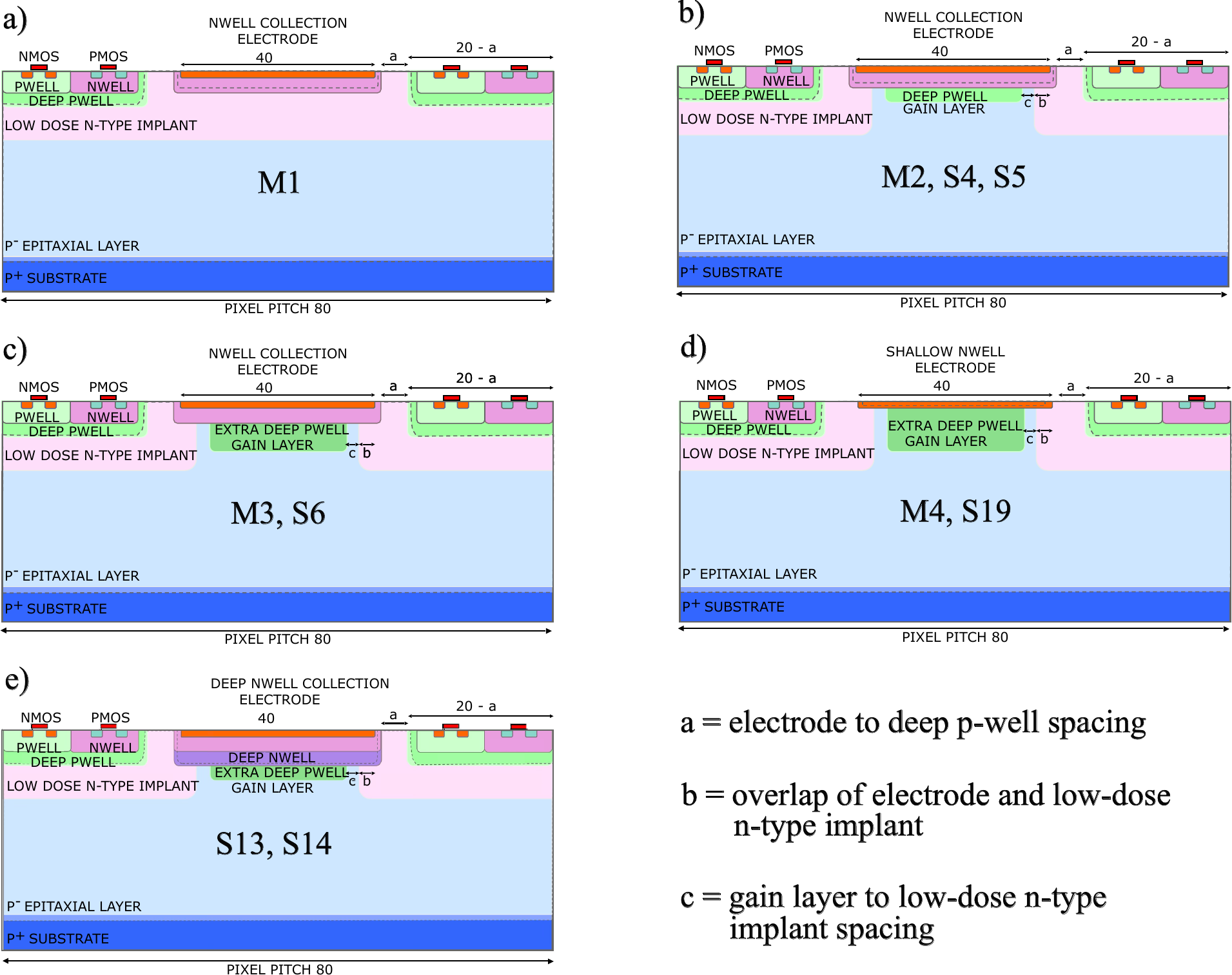}

    \caption{Cross section of CASSIA sensors: Sensor M1 (a) with standard n$^{+}$-electrode without gain layer, Sensor M2/S4/S5 (b) with a deep p-well (DP) gain layer, Sensor M3/S6 (c) with a extra deep p-well (XDP) gain layer, Sensor M4/S19 (d) with a extra deep p-well gain layer and shallow n$^{+}$-electrode implant and Sensor S13/S4 (e) with a deep n$^{+}$-electrode and extra deep p-well (XDP) gain layer.
    }
    \label{fig:implants}
\end{figure}

The CASSIA sensor is implemented in the 180\,nm TowerSemiconductor imaging process modified for MALTA fabrication \cite{Pernegger_2023,Snoeys:2017hjn,Munker:2019vdo,PERNEGGER2021164381}. The sensitive region is a 30$\,\rm\upmu$m thick, depleted p-type epitaxial layer on top of a p+ substrate, as shown in Figure~\ref{fig:implants}. 

The readout electronics is implemented within a deep p-well region around the pixel electrode. The deep p-well extends underneath the n-wells of pMOS transistors, as seen in Figure~\ref{fig:implants}, preventing them from collecting ionization electrons generated in the epitaxial layer, ensuring all generated electrons are collected only by the central n-electrode. From here on we refer to this deep p-well region as the electronics deep p-well. The collection electrode is made as a circular n$^{+}$-implant with a radius of 20$\,\rm\upmu m$, which is from here on referred to as the n$^{+}$-electrode. 

The pixel implant geometry used in the 180~nm TowerSemiconductor process has been optimized for fast charge collection and high radiation hardness in collaboration with the foundry. Figure~\ref{fig:implants} shows the cross-sections of different implant designs. The ionisation signal is collected by the n$^{+}$-electrode in the pixel center, which is connected to an external pad for bias supply and signal readout. The n$^{+}$-electrode is biased to a positive voltage to deplete the sensor bulk. The n$^{+}$-electrode is surrounded by the electronics deep p-well at a certain distance, which can be used for analog and digital circuitry in future sensors.  The substrate and electronics deep p-well are kept at ground for the measurements shown in this paper. Beneath the n$^{+}$-electrode, the gain layer is implanted. We used two different depths, referred to as "deep p-well (DP) gain layer" and "extra deep p-well (XDP) gain layer". These different gain layer profiles allow us to investigate their impact on gain and breakdown behavior. The main goal in the design of the different CASSIA versions is to obtain a uniform lateral distribution of avalanche multiplication in the active region between the gain layer and the collection electrode, avoiding breakdown at the collection electrode edge and covering as much pixel area as possible, all while achieving full depletion of the epitaxial layer. 

The standard 180\,nm TowerSemiconductor imaging process is supplemented with a low-dose $n$-type implant across the full pixel matrix \cite{Snoeys:2017hjn,PERNEGGER2021164381}. This low-doped n-well layer shifts the pn-junction from the electrode to the bulk of the sensor, which helps to deplete the silicon sensor and electrically separates the electronics deep p-well from the p-type substrate. 
The substrate is either p$^{-}$ epitaxial layer ($>$1000$\,\rm\Omega$cm) or the high-resistivity Czochralski bulk silicon ($>$800$\,\rm\Omega$cm) without epitaxial growth\footnote{As given in substrate manufacturing specification}. The first CASSIA sensors (CASSIA-1) have been manufactured on both substrate types. The substrate bias is supplied in parallel from the top side through a substrate p-type connection as well as the sensor backside through an electrical connection on the chip-carrier PCB. A positive bias voltage of up to 160\,V is applied to the n$^{+}$-electrode. The voltage difference between the positively biased electrode and the grounded substrate generates an electric field high enough to lead to impact ionization in the gain layer region, resulting in internal charge multiplication.  The low-dose n-type implant is depleted from its junctions to the electronics deep p-well on one side and the p-type substrate on the other side. The choice of gain layer doping concentration, the distance between the gain layer and the n$^{+}$-electrode, and the size of the gain layer with respect to the n$^{+}$-electrode influence the detector operation, breakdown voltage, and observed gain.

\begin{figure}
    \centering
    \includegraphics[width=0.5\textwidth]{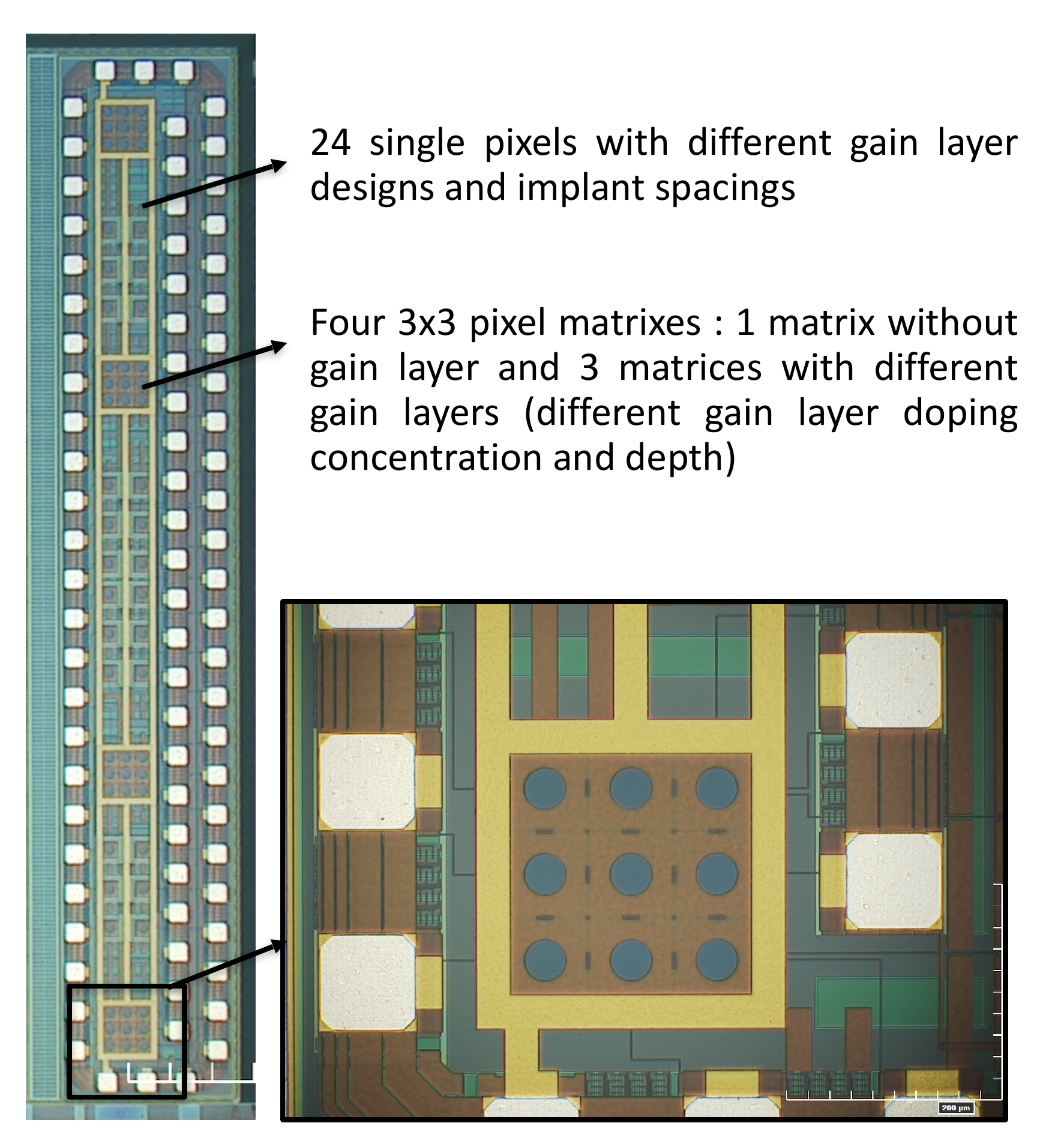}
     \caption{The CASSIA-1 chip, containing 24 single pixels and four 3x3 matrices.}
    \label{fig:CASSIA}
\end{figure}

The CASSIA-1 prototype sensor is 1$\times$5~mm$^{2}$ and contains four 3$\times$3 pixel matrices, (denoted "M") and 24 individual pixels (denoted "S"), as shown in Figure~\ref{fig:CASSIA}. All pixel signals from matrices and single pixels are routed directly to wirebond pads on the die periphery. 
The 3$\times$3 pixel matrices and single pixels feature different designs, based on three electrode implants, shallow, standard and deep n$^{+}$-electrode, and two gain layer implants, the DP and XDP gain layer p-implants. 
The top of each electrode is kept free of metallisation to enable laser testing. While the n$^{+}$-electrodes are always circular with  40$\,\rm\upmu$m diameter, the gain layer radius and its distance to electronics deep p-well are varied. One should note the diameter of the collection electrode has been increased by an order of magnitude compared with the MALTA development to achieve uniform gain over the gain layer area, leading to an increase in sensor capacitance, which is estimated to be about a factor of 10. However, the CASSIA pixels can easily achieve a sensor gain well over 10, as seen later on, meaning that the ratio of collected charge over sensor capacitance should improve by implementing internal sensor gain. The CASSIA-1 prototypes were thinned to 300$\,\rm\upmu$m thickness without back-side implantation or metallization. 

The four 3$\times$3 pixel matrices and seven single pixels presented in this paper, shown in Figure~\ref{fig:implants}, consist of the following designs:
\begin{enumerate}
\item Matrix M1 pixels (Figure~\ref{fig:implants}a) have n$^{+}$-electrodes surrounded by an electronics deep p-well. The pixels do not contain a gain layer. This matrix serves as a reference to verify the electrical behavior without gain layer (e.g. for breakdown behavior).
\item Matrix M2 pixels (Figure~\ref{fig:implants}b) have n$^{+}$-electrodes surrounded by an electronics deep p-well. The gain layer is formed by a circular deep p-well (DP) in the center under the n$^{+}$-electrode with a diameter of 20$\,\rm\upmu$m. Single pixel S4 has the same implant configuration and gain layer diameter, but the spacing between n$^{+}$-electrode and electronics deep p-well is reduced. Single pixel S5 has the identical design to M2 but with a gain layer diameter of 28$\,\rm\upmu$m.
\item Matrix M3 pixels (Figure~\ref{fig:implants}c) have n$^{+}$-electrodes surrounded by an electronics deep p-well. The gain layer is formed by a circular extra deep p-well (XDP) in the center under the n$^{+}$-electrode with a diameter of 12$\,\rm\upmu$m. Single pixel S6 has an identical design to M3 but with a gain layer diameter of 28$\,\rm\upmu$m.
\item Matrix M4 pixels (Figure~\ref{fig:implants}d) have shallow n$^{+}$-electrodes surrounded by an electronics deep p-well. The gain layer is formed by a circular extra deep p-well (XDP) in the center under the electrode with a diameter of 12$\,\rm\upmu$m. Single pixel S19 has an identical implant design to M4 but with a gain layer diameter of 28$\,\rm\upmu$m.
\item Single pixels S13 and S14 (Figure~\ref{fig:implants}e) have a deep n$^{+}$-electrodes surrounded by an electronics deep p-well. The gain layer is formed by a circular extra deep p-well (XDP) in the center under the electrode with a diameter of 20$\,\rm\upmu$m for S13, and 28$\,\rm\upmu$m for S14.

\end{enumerate}

\section{Simulation of CASSIA field configuration and $I-V$ curves} \label{sec:sim}

To evaluate the expected behavior of different gain layer designs we carried out TCAD simulations\footnote{Synopsys Sentaurus Device User Guide, S-2021.06-SP1} to study breakdown behavior (e.g. breakdown voltages) as a function of dimensions and implant configurations. The implantation profiles used for simulations are based on two sets of profiles: (a) foundry provided process simulations  and (b) generic profiles adjusted to match simulation to data. The two profile sets are not related and were established and applied independently from each other. For all TCAD simulations, the biasing conditions match those used during the measurements, i.e. sensor backside/substrate and p-well are grounded, n$^{+}$-electrode biased to positive voltage. Impact ionization and breakdown are modeled using the Okuto-Crowell model \cite{OKUTO1975161}. 

\begin{figure}
    \centering
    \includegraphics[width=0.7\textwidth]{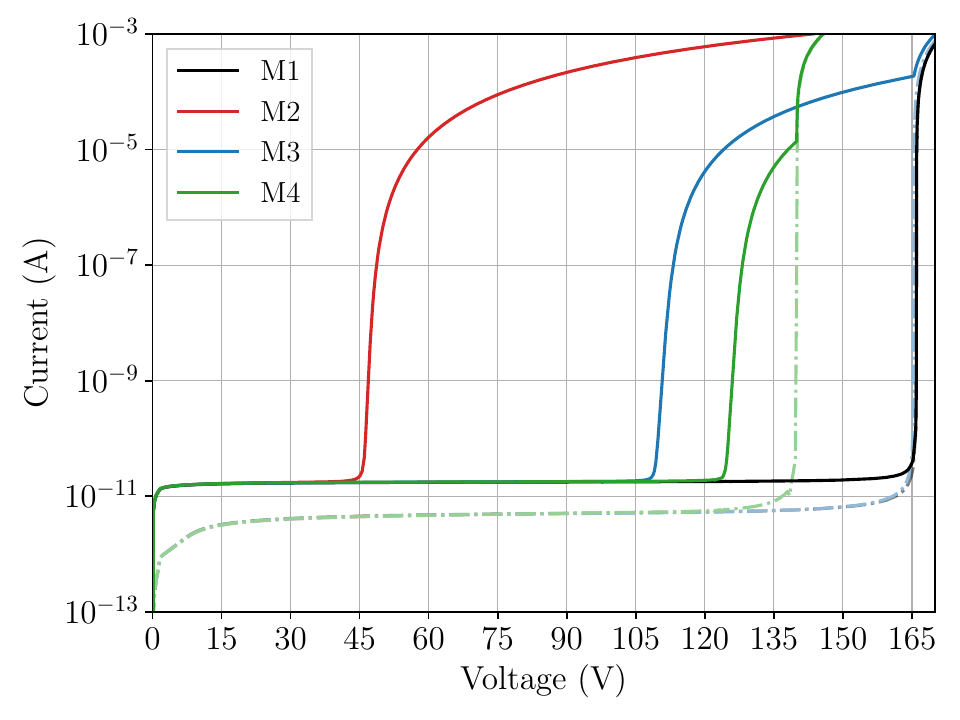}
    \caption{TCAD simulated $I-V$ curve for matrix M1 (black) without gain layer, M2 (red) with a deep p-well (DP) gain layer, M3 (blue) with a extra deep p-well (XDP) gain layer, M4 (green) with a extra deep p-well (XDP) gain layer and shallow n$^{+}$-electrode. Solid curves show current from the n$^{+}$-electrode to substrate; dashed curves show current from the n$^{+}$-electrode to the electronics deep p-well.
    }
    \label{fig:IVsim}
\end{figure}

Figure~\ref{fig:IVsim} shows simulated $I-V$ curves from 2D-cylindrical TCAD simulations using foundry-provided profiles for matrices M1 (black), M2 (red), M3 (blue) and M4 (green) on epitaxial layer sensors. Solid curves show the total current through the n$^{+}$-electrode; dashed curves shows the current between the n$^{+}$-electrode and the electronics deep p-well. For matrix M1 (without gain layer), we observe no current increase with voltage until breakdown occurs between the n$^{+}$-electrode and electronics deep p-well at 165\,V. Matrices with gain layer (M2, M3, M4) have a high electric field in between n$^{+}$-electrode and gain layer which enables impact ionization and charge multiplication, which leads to break down between the n$^{+}$-electrode and the substrate via the gain layer. Changes in gain layer implant configuration result in increasing breakdown voltages: M2 (47V), M3 (110V), and M4 (126V). Matrix M2 has shallower gain layer implantation than M3/M4 which may explain this qualitative behavior. The gain layer implants for M3 and M4 are identical, however M4 has a shallower n$^{+}$-electrode implantation compared to M3, which reduces the electric field in the multiplication region and leads to a higher breakdown voltage. Amplification continues to increase with voltage until the n$^{+}$-electrode breaks through to the p-well: at 165\,V for M3 and 142\,V for M4. The TCAD simulation reproduces the behavior of different sensors' $I-V$ curves, and the simulated breakdown voltages match well with measured values (see Section \ref{sec:IV}).

\begin{figure}[t!]
\centering
\includegraphics[width=0.85\textwidth]{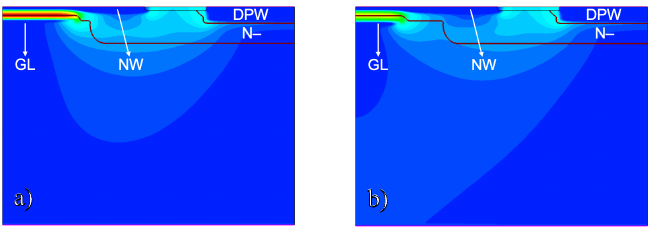}
\caption{Right half of CASSIA structure showing generic TCAD simulations of electric field distribution at 55\,V: (a) Matrix M2 with deep p-well (DP) gain layer of 20$\,\rm\upmu$m diameter and (b) Matrix M3 with extra deep p-well (XDP) gain layer of 12$\,\rm\upmu$m diameter. The colour scale in arbitrary units is the same for both graphs.}
\label{tcad_field}
\end{figure}

\begin{figure}[t!]
\centering
\includegraphics[width=0.85\textwidth]{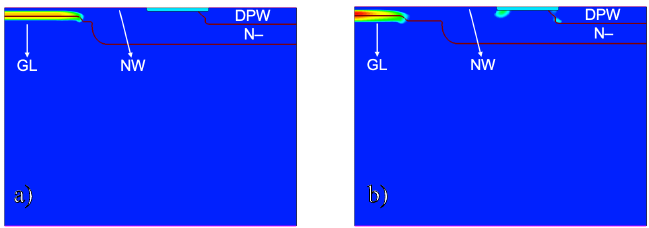}
\caption{Generic TCAD simulations of avalanche multiplication (impact ionization) rate distribution in CASSIA with 8$\,\rm\upmu$m n$^{+}$-electrode to electronics deep p-well spacing: (a) Matrix M2 with DP gain layer biased at 56\,V, (b) Matrix M3 with XDP gain layer biased at 103\,V. The colour scale in arbitrary units is the same for both graphs.}
\label{tcad_impact1}
\end{figure}

\begin{figure}[t!]
\centering
\includegraphics[width=0.85\textwidth]{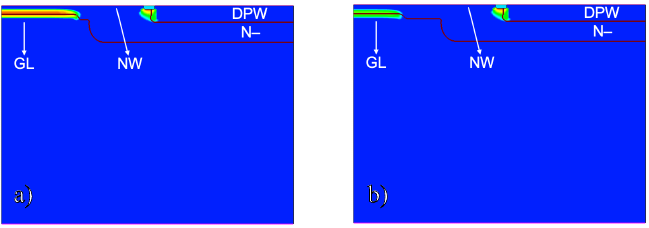}
\caption{Generic TCAD simulations of avalanche multiplication (impact ionization) rate distribution in CASSIA at 55\,V with the n$^{+}$-electrode to electronics p-well distance intentionally reduced to 1$\,\rm\upmu$m: (a) Matrix M2 with DP gain layer, (b) Matrix M3 with XDP gain layer. The colour scale in arbitrary units is the same for both graphs.}
\label{tcad_impact2}
\end{figure}

\begin{figure}[t!]
\centering
\includegraphics[width=0.85\textwidth]{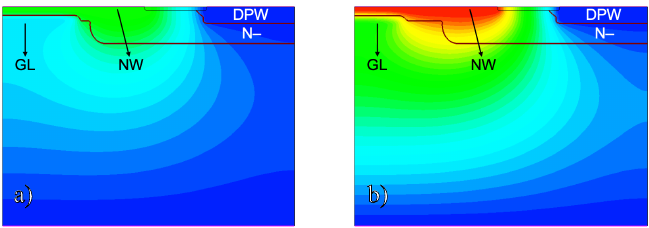}
\caption{Generic TCAD simulations of electrostatic potential distribution in CASSIA with 8$\,\rm\upmu$m n$^{+}$-electrode to electronics deep p-well spacing: (a) Matrix M2 with DP gain layer biased at 56\,V, (b) with Matrix M3 with XDP gain layer biased at 103\,V. The colour scale in arbitrary units is the same for both graphs.}
\label{tcad_potential}
\end{figure}

In a second set of TCAD simulations using generic profiles, we focused on the internal distribution of critical electrical quantities and their correlation with measurements presented in Section~\ref{dc_measurements}. The doping profiles of DP regions for matrix M2 and XDP regions for matrix M3 were adjusted as generic profiles in 2D-cylindrical TCAD simulations until the measured breakdown voltages are reproduced. Figure~\ref{tcad_field} shows the electric field distribution in both structures at 55\,V. For clarity the gain layer area is indicated as "GL", the electrode n-well is indicated with "NW", the electronics deep p-well is indicated as "DPW" and the low doped n$^{-}$-layer is indicated with "N-". Note that results in  Figure~\ref{tcad_field} use adjusted generic profiles rather than process simulated profiles. The electric field peaks in the active region between the n$^{+}$-electrode and DP/XDP gain layers, while the electric field at the periphery is lower. This confirms that the breakdown occurs in the active region and is successfully prevented at the periphery. 

The peak electric field is higher in the DP gain layer structure than the XDP gain layer structure, resulting in a lower breakdown voltage. Figure~\ref{tcad_impact1} shows the avalanche multiplication (impact ionization) rate distribution in both structures at 5\,V below breakdown. Avalanche multiplication is maximal in the active region for both structures. This indicates that the 8$\,\rm\upmu$m distance between the n$^{+}$-electrode and electronics deep p-well region is large enough to accommodate the voltage drop from electrode bias; more than 56\,V for matrix M2 with DP gain layer and 103\,V for matrix M3 with XDP gain layer. Figure~\ref{tcad_impact2} shows avalanche multiplication distribution for a reduced n$^{+}$-electrode to electronics deep p-well distance of 1$\,\rm\upmu$m. In this case, the peak rate occurs at the periphery rather than under the n$^{+}$-electrode, failing to amplify signals generated in the active region and resulting in breakdown towards the electronics deep p-well at around 55\,V for both structures. Therefore, sufficient distance between the n$^{+}$-electrode and electronics deep p-well is critical for achieving avalanche multiplication in the active region.

In the actual chip design, the electric field at the periphery is also influenced by the surface conditions such as interface and oxide charges, metallization, defects, and contamination. These surface conditions are technology specific and depend on stress from the applied voltage, environmental radiation, temperature, and operating history. The worst-case surface conditions must be considered for a particular application, potentially requiring a larger n$^{+}$-electrode to electronics deep p-well distance than predicted in ideal TCAD simulations. Figure~\ref{tcad_potential} shows the potential distribution in both CASSIA sensors M2 and M3 at 5\,V below breakdown. The DP gain layer in M2 is at a potential of around 20\,V, meaning that this voltage is dropped across the epitaxial layer, whereas the remaining 36\,V is dropped between the electrode and the DP gain layer. Similarly, the XDP gain layer in M3 is at 60\,V, corresponding to the voltage drop across the epitaxial layer, with the remaining 53\,V dropped in the gain layer region. In both cases, the epitaxial layer is fully depleted due to the low-doped n-well layer and could act as a photo/particle sensitive region from which electrons drift towards the electrode.

\section{Electrical and Optical Characterization}

In the following section, we summarize results of the first measurements on the CASSIA sensor. The sensors were tested in laboratory in a light-shielded enclosure. The sensor die is mounted to a PCB which supplies the sensor with operating voltages for the electronics deep p-well, substrate and a bias to n$^{+}$-electrode. Alternatively bare dies are probed on a probestation. Each pixel signal is connected to a bond pad at the die's periphery and via the PCB to an external amplifier for pulse measurements or to a source-measure unit (SMU) for static DC current measurements.

During the measurements, the substrate and electronics deep p-well are kept at ground potential while the n$^{+}$-electrode is biased at positive voltage. In the 3x3 pixel the central pixel is connected to either SMU for DC measurements or external amplifier for pulse measurements.

Break down behavior for different CASSIAs structures will be analyzed in the following sections, and for the purpose of describing the breakdown behavior we define two transition voltages: the first transition voltage is the voltage at which first amplification is observed. We define this as $V_{\text{LGAD}}$, the voltage at which the $I-V$ curve shows the transition to low-gain avalanche mode. The second transition voltage $V_{\text{BR}}$ is the voltage at which hard breakdown occurs in the $I-V$ curve between electrode and gain layer, corresponding to the transition from LGAD to SPAD amplification mode. For the determination of $V_{\text{LGAD}}$ and $V_{\text{BR}}$ based on the measured $I-V$  curves we apply the logarithmic derivative $LD=\frac{d \log{I}}{dV}$ as well as the inverse logarithmic derivative $ILD=1/LD$. The value for $V_{\text{LGAD}}$, respectively $V_{\text{BR}}$, is the mean of maximum in $LD$ and minimum in $ILD$ \cite{KLANNER201936}.

\subsection{Electrical and optical characteristics in DC measurements} \label{dc_measurements}

\begin{figure}[t!]
\centering
\begin{subfigure}{0.45\textwidth}
\includegraphics[width=0.9\textwidth]{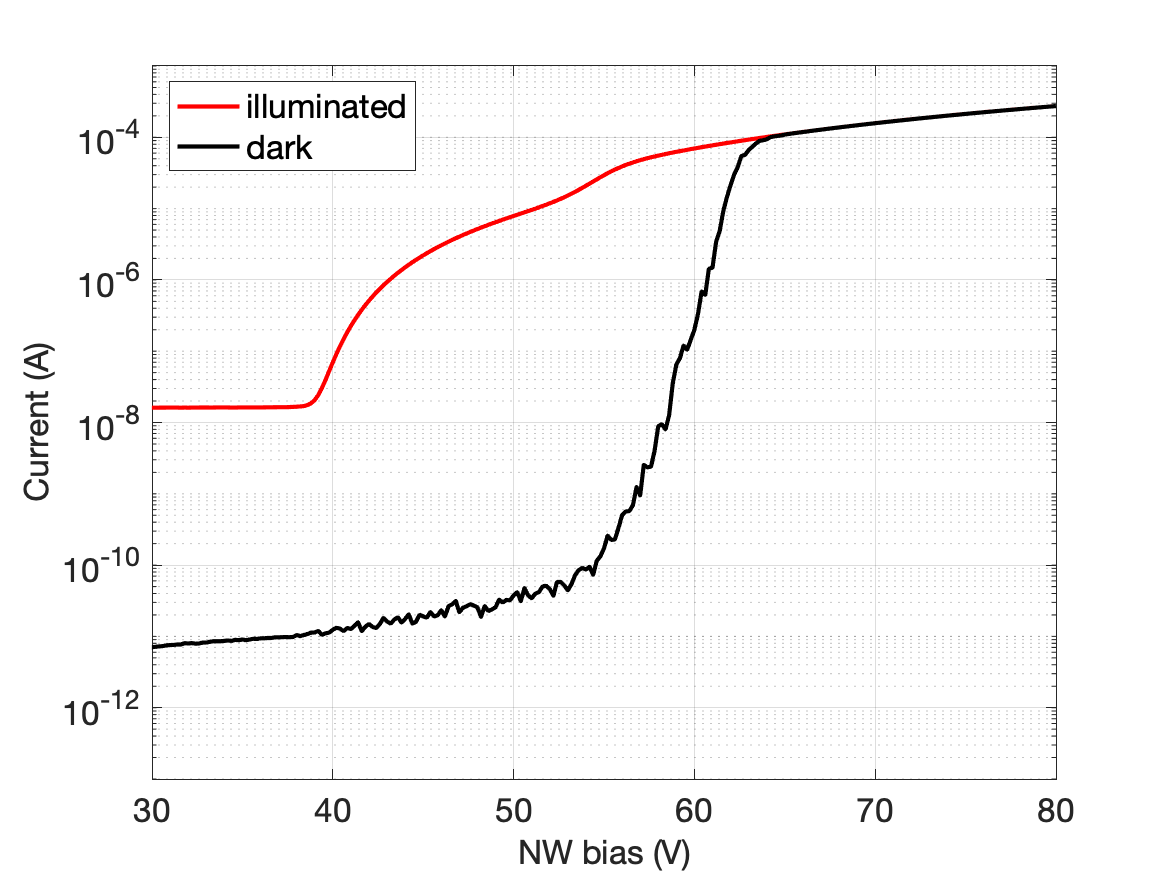}
\caption{}
\end{subfigure}
\begin{subfigure}{0.45\textwidth}
\includegraphics[width=0.9\textwidth]{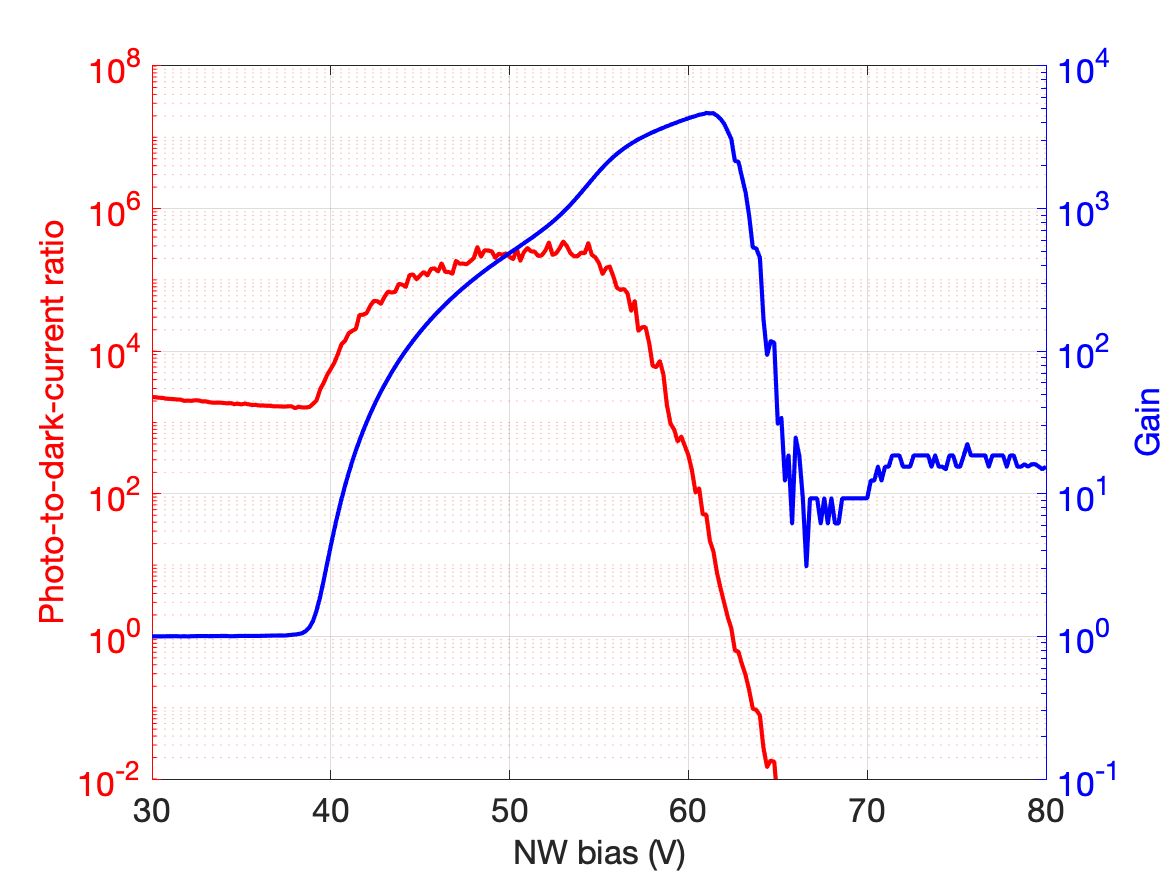}
\caption{}
\end{subfigure}
\caption{Measured current-voltage ($I-V$) characteristics of  CASSIA matrix M2 (DP gain layer) with the n$^{+}$-electrode connected to positive bias and with the substrate and electronics deep p-well region grounded, (a) dark and photo currents, (b) extracted photo-to-dark-current ratio and gain calculated as the ratio between the photocurrent at a given reverse voltage and photocurrent at 30\,V. The substrate contact is implemented at the top surface of the sensor at the pixel periphery.}
\label{iv_m2}
\end{figure}

The measured reverse $I-V$ characteristics of CASSIA matrix M2 (DP gain layer) in dark conditions and under continuous visible light illumination are shown in Figure~\ref{iv_m2}a. During the measurements, the DC currents of substrate, electronics deep p-well and n$^{+}$-electrode are continuously measured. The hard breakdown voltage, $V_{\text{BR}}$, equals 54.4\,V. The dark current is under 10\,pA at reverse voltages under 30\,V and is limited by the noise of the measurement setup. The current increase with voltage approaches linear growth at voltages above 63\,V is due to a series resistance, mostly contributed by the epitaxial layer resistance under the substrate contact at the periphery. The photocurrent, defined as the difference between the illuminated current and dark current, is constant with voltage up to 38\,V, which implies that the photosensitive region does not increase with voltage in this range. Above 38\,V, the photocurrent increases, which is caused by the onset of avalanche multiplication. Such an increase is not observed in the dark current characteristics, since it is under the noise floor. The calculated photo-to-dark-current ratio (PDCR$=\frac{I_{\text{illuminated}}-I_{\text{dark}}}{I_{\text{dark}}}$, calculated as the difference between the illuminated and dark current divided by the dark current itself) and gain, defined as a ratio of photocurrent in each bias point and photocurrent at low voltage (at 30 V) before the onset of avalanche multiplication, are shown in figure~\ref{iv_m2}b. The gain becomes higher than 1 at 38\,V and gradually increases to above 5000 at 62\,V, where its further increase is limited by series resistance. Such a voltage range of 24\,V with gain increase is useful both for LGAD and SPAD mode of operation, where gain can be precisely controlled by adjusting the bias point.

\begin{figure}[t!]
\centering
\begin{subfigure}{0.45\textwidth}
\includegraphics[width=0.9\textwidth]{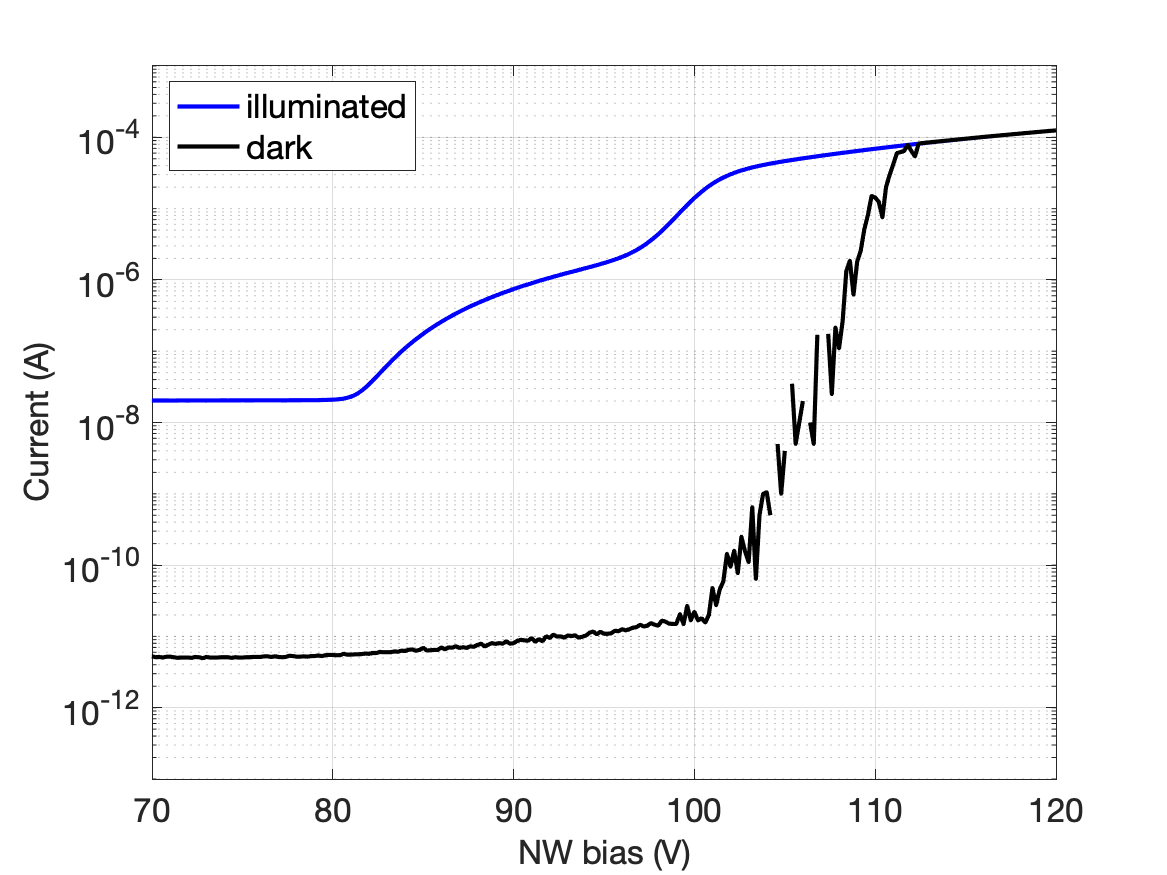}
\caption{}
\end{subfigure}
\begin{subfigure}{0.45\textwidth}
\includegraphics[width=0.9\textwidth]{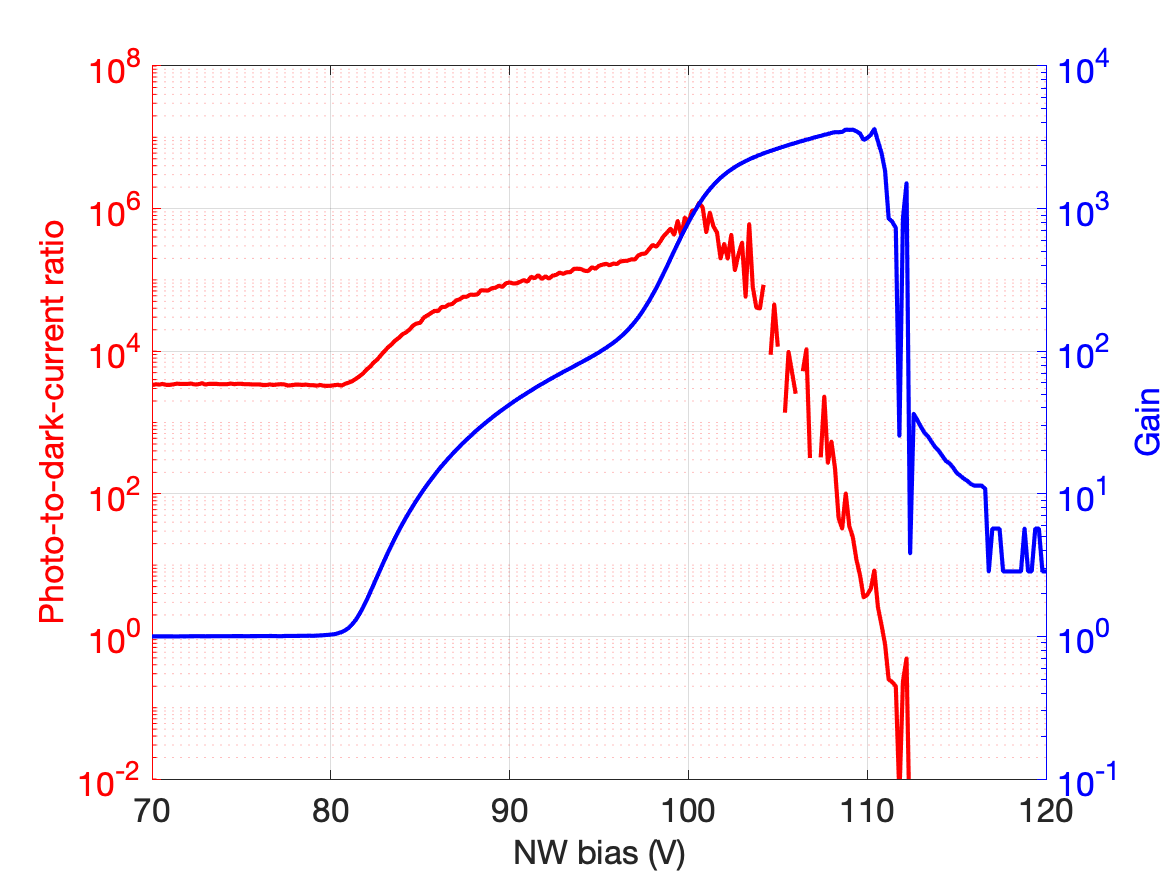}
\caption{}
\end{subfigure}
\caption{Measured current-voltage ($I-V$) characteristics of CASSIA matrix M3 (XDP gain layer)  with the n$^{+}$-electrode connected to positive bias and with the substrate and electronics deep p-well region grounded, (a) dark and photo currents, (b) extracted photo-to-dark-current ratio and gain calculated as the ratio between the photocurrent at a given reverse voltage and photocurrent at 30\,V.}
\label{iv_m3}
\end{figure}

The measured $I-V$  characteristics of CASSIA matrix M3 (XDP gain layer)  pixel is shown in Fig.~\ref{iv_m3}a. The hard breakdown voltage, $V_{\text{BR}}$, equals 99.0\,V. A higher breakdown voltage compared to M2 is a result of its deeper gain layer implantation and consequently reduced charge at the junction between the n$^{+}$-electrode and the XDP gain layer, and decreased electric field peak in the avalanche multiplication region. The photocurrent shows an amplification from 82\,V onwards. The resulting photo-to-dark-current ratio (PDCR) reaches a peak around $10^{6}$ at 100\,V, which is around 3 times higher than in M2 with DP gain layer under the same illumination conditions. The CASSIA matrix M3 exhibits a gain higher than 1 over the range of 30\,V (from 82\,V to 112\,V), reaching a value above 4000 at 110\,V, where its further increase is limited by the series resistance.

\begin{figure}
\centering
\includegraphics[width=0.7\textwidth]{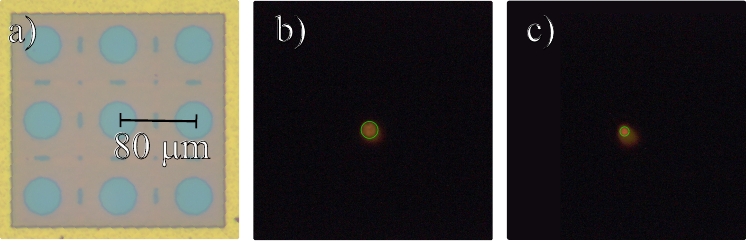}
\caption{a) Top-view optical micrograph of a 3$\times$3 CASSIA pixel matrix with 80\,$\upmu$m pixel pitch, b) light emission from M2 with DP gain layer (20\,$\upmu$m diameter) where the central pixel is biased at 94\,V, c) light emission from M3 with XDP gain layer (12\,$\upmu$m diameter) where the central pixel is biased at 160\,V. Green circles in b) and c) indicate gain layer boundaries; all micrographs are shown at the same magnification.}
\label{emission}
\end{figure}

The distance between the n$^{+}$-electrode and electronics deep p-well is the same in both CASSIA M2 and M3. As those two versions of CASSIA have different breakdown voltages, it means that the avalanche multiplication process is dominant in the active region and the breakdown mechanism is not initiated at the periphery, between the n$^{+}$-electrode and the electronics deep p-well. In order to verify the avalanche multiplication distribution, light emission tests were conducted on both CASSIA matrix versions. The light emission distribution is uniform in both CASSIA M2 and M3 versions biased well above the breakdown voltage, without a higher intensity emission at periphery, as shown in Figure~\ref{emission}. CASSIA M2 has a larger diameter of the DP gain layer (20$\,\rm\upmu$m) than CASSIA M3 with an XDP gain layer diameter of 12$\,\rm\upmu$m, which both correspond to the measured diameter of light emission.

\subsection{Position dependent optical illumination response}

\begin{figure}[t!]
\centering
\includegraphics[width=0.7 \textwidth]{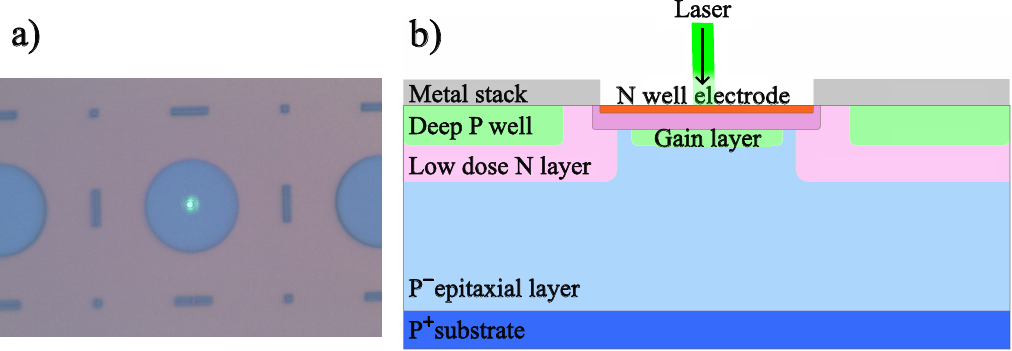}
\caption{a) Top-view micrograph of a CASSIA pixel with 532\,nm laser focused at the pixel center (80\,$\upmu$m pitch), b) cross-sectional schematic showing the layer structure with laser beam path. The diagram shows the n$^{+}$-electrode electrode (N well electrode), electronics deep p-well (Deep P well), gain layer, low-dose n-type implant, epitaxial layer, and substrate. }
\label{scan_setup}
\end{figure}

The position-dependent photo response of CASSIA pixels has been tested by continuous-wave, focused laser beam illumination. The sensors have been illuminated from the top side by a laser with wavelengths of 532\,nm and 785\,nm. The beam with a full-width-half-maximum (FWHM) of 2$\,\rm\upmu$m is scanned across the sensor with a step of 1$\,\rm\upmu$m. Such FWHM and scanning resolution make it possible to have the absorption of the complete laser energy either in the active region or at the periphery and to examine the transition region, as can be seen in Figure~\ref{scan_setup}a showing the top-view of CASSIA illuminated with the 532\,nm laser beam. The sample is mounted on an X-Y-Z stage which allows for precision scanning of the laser beam across the pixel. The pixel area outside the n$^{+}$-electrode is covered by metal layers, as sketched in Figure~\ref{scan_setup}b.

\begin{figure}[t!]
\centering
\begin{subfigure}{0.45\textwidth}
\includegraphics[width=\textwidth]{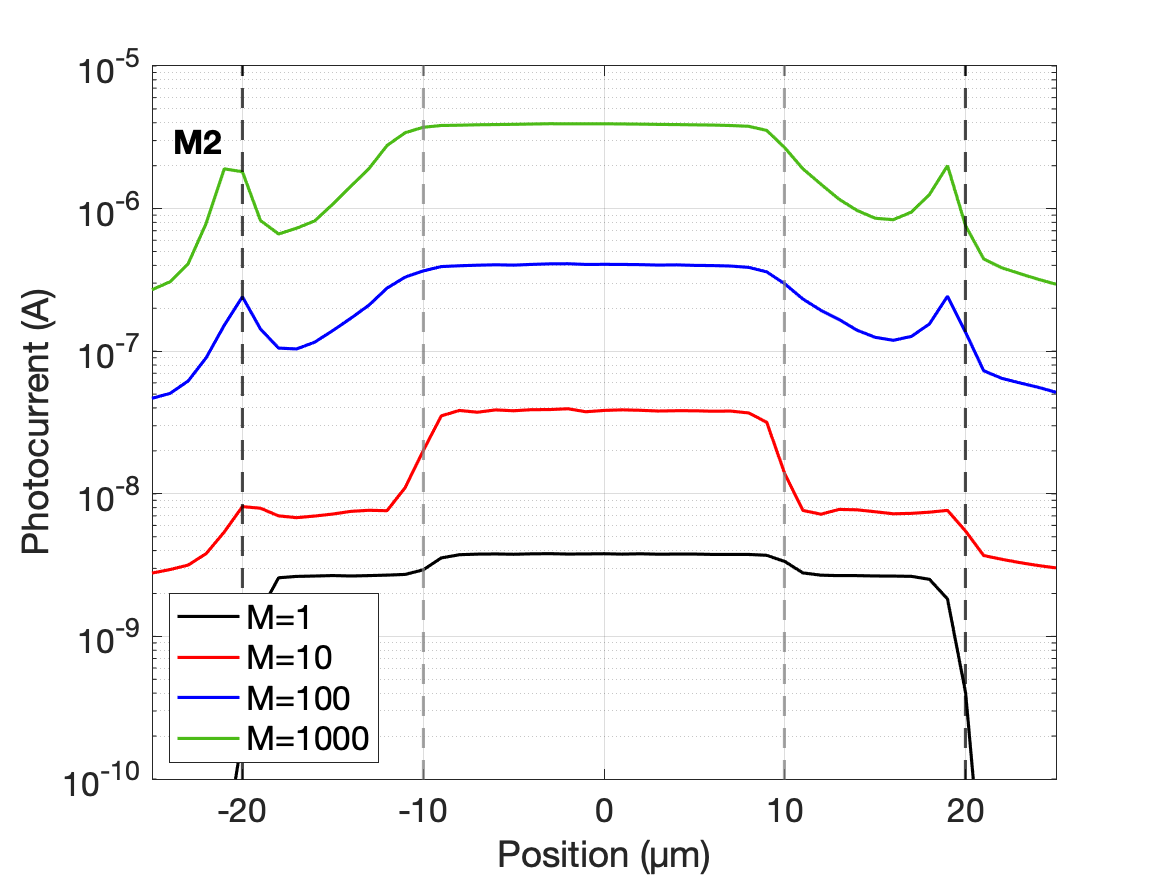}
\caption{}
\end{subfigure}
\begin{subfigure}{0.45\textwidth}
\includegraphics[width=\textwidth]{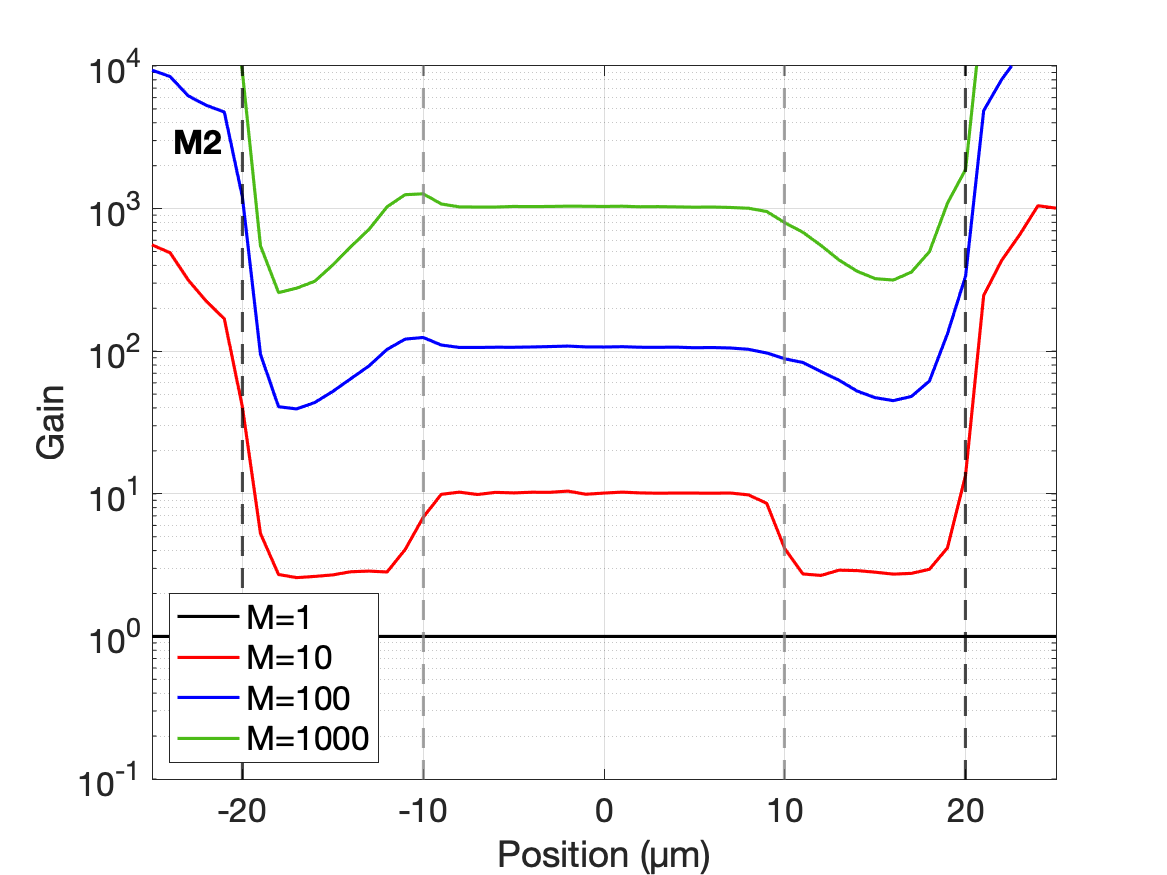}
\caption{}
\end{subfigure}
\caption{Measured photoresponse of CASSIA M2 (DP gain layer) with an n$^{+}$-electrode radius of 20 um and DP gain layer radius of 10$\,\rm\upmu$m scanned by a laser beam with a FWHM of 2$\,\rm\upmu$m and 532\,nm wavelength, biased at reverse voltages of 30\,V (black curve), 39.1\,V (red curve), 41.1\,V (blue curve) and 44.8\,V (green curve) corresponding to gains of 1, 10, 100, and 1000, respectively. (a) Dependence of photocurrent on radial beam position and (b) gain calculated as the measured photocurrent at the given voltages divided by the photocurrent without gain (black curve).}
\label{scan_m2_532}
\end{figure}

Photocurrents of CASSIA have been measured at bias voltages that were adjusted to achieve different gains. The laser power was set to obtain a photocurrent in the nanoampere range in the middle of the pixel, well above the noise floor, even at low bias voltages. Figure~\ref{scan_m2_532} shows the measured photocurrent dependence on laser beam position for CASSIA M2, with a laser wavelength of 532\,nm. The photocurrent is first measured at a voltage of 30\,V, which is before the onset of avalanche multiplication, i.e. without gain, and then at voltages of 39.1\,V, 41.1\,V and 44.8\,V, corresponding to the gains of 10, 100 and 1000, respectively, with the laser beam in the middle of the sensor. The spatial distribution of gains, calculated as ratio between the measured photocurrent at given voltage to a photocurrent without avalanche multiplication at 30\,V, are shown in Figure~\ref{scan_m2_532}b. The photocurrent is constant across the active region, above the DP gain layer with radius of 10$\,\rm\upmu$m, without any spikes around the DP gain layer region edges that could come from peripheral avalanche multiplication. Local spikes at the n$^{+}$-electrode edge, around a radius of 20$\,\rm\upmu$m, are potentially caused by light reflection/diffraction into the active area from the edges of the metal above the sensor (Figure~\ref{scan_setup}). In any case, the photocurrent from the illumination at 20$\,\rm\upmu$m is lower than the photocurrent from illumination in the active region for all bias voltages. 

The photocurrent, and consequently the gain, drop gradually when moving away from the gain layer region, i.e. for radii greater than 10$\,\rm\upmu$m, which implies that even photons impinging the sensor outside the DP gain layer would still result in an amplified photocurrent. For example, at a bias of 41.1\,V resulting in an active-region gain of 100, the photons impinging the sensor 5$\,\rm\upmu$m away from the active region still experience a gain above 40, which could increase the fill factor with gains above a certain threshold beyond the mere area of the avalanche multiplication layer. High gains under the metal layer, at radii above 20$\,\rm\upmu$m, are just a consequence of low photocurrent for a gain of 1 in the denominator.

\begin{figure}[t!]
\centering
\begin{subfigure}{0.45\textwidth}
\includegraphics[width=\textwidth]{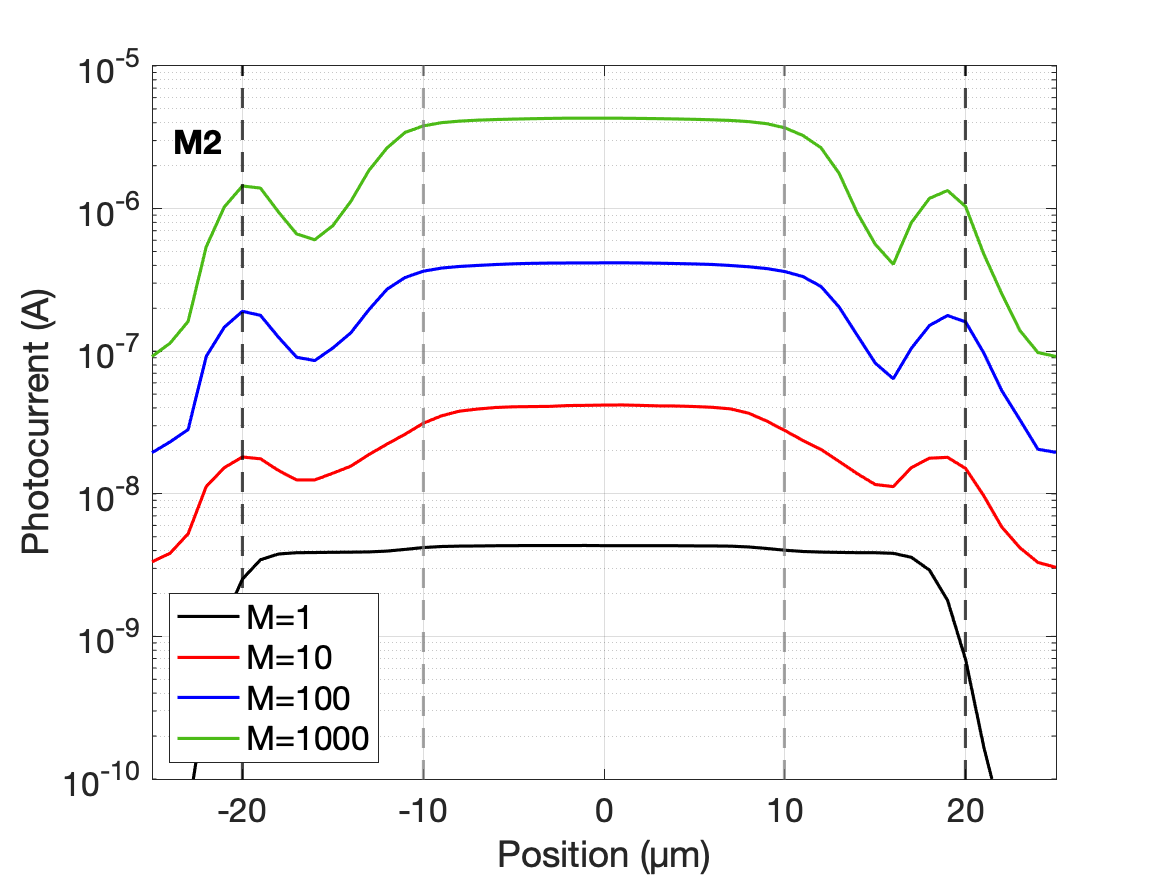}
\caption{}
\end{subfigure}
\begin{subfigure}{0.45\textwidth}
\includegraphics[width=\textwidth]{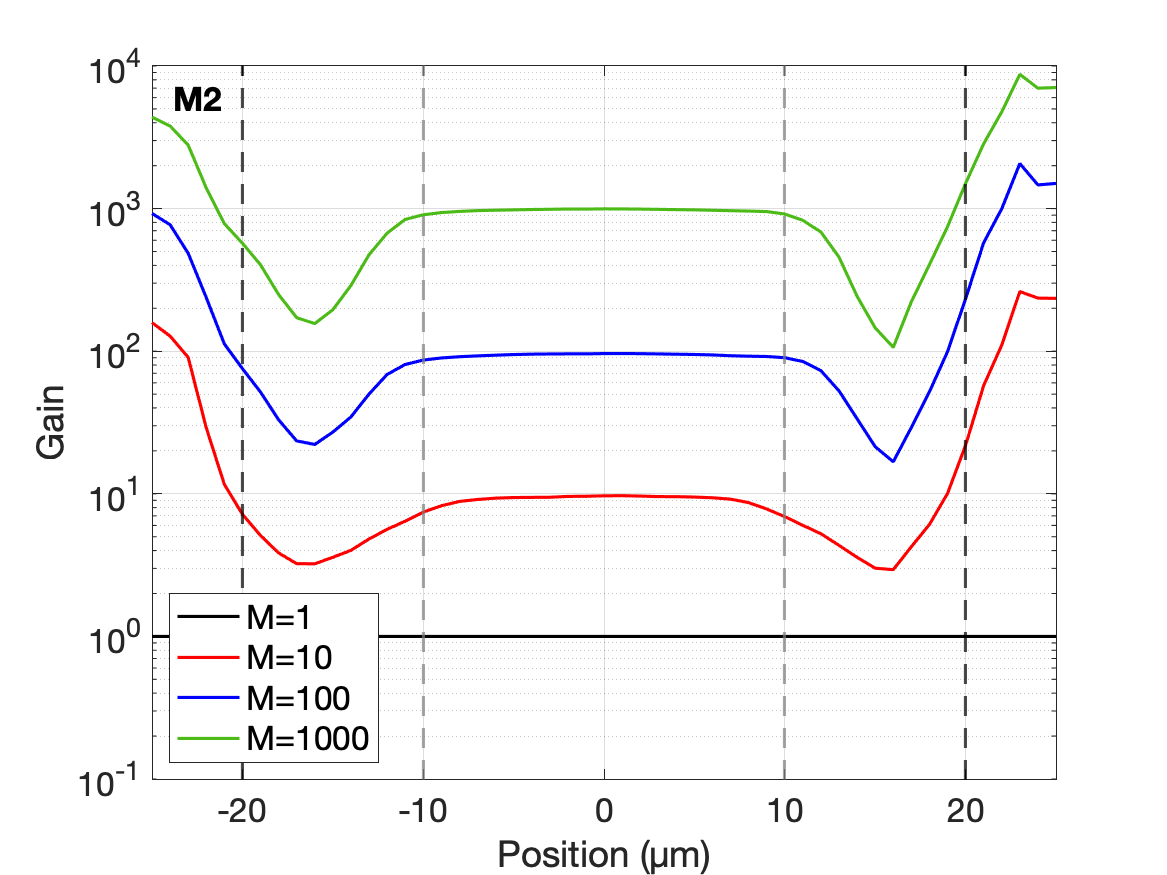}
\caption{}
\end{subfigure}
\begin{subfigure}{0.45\textwidth}
\includegraphics[width=\textwidth]{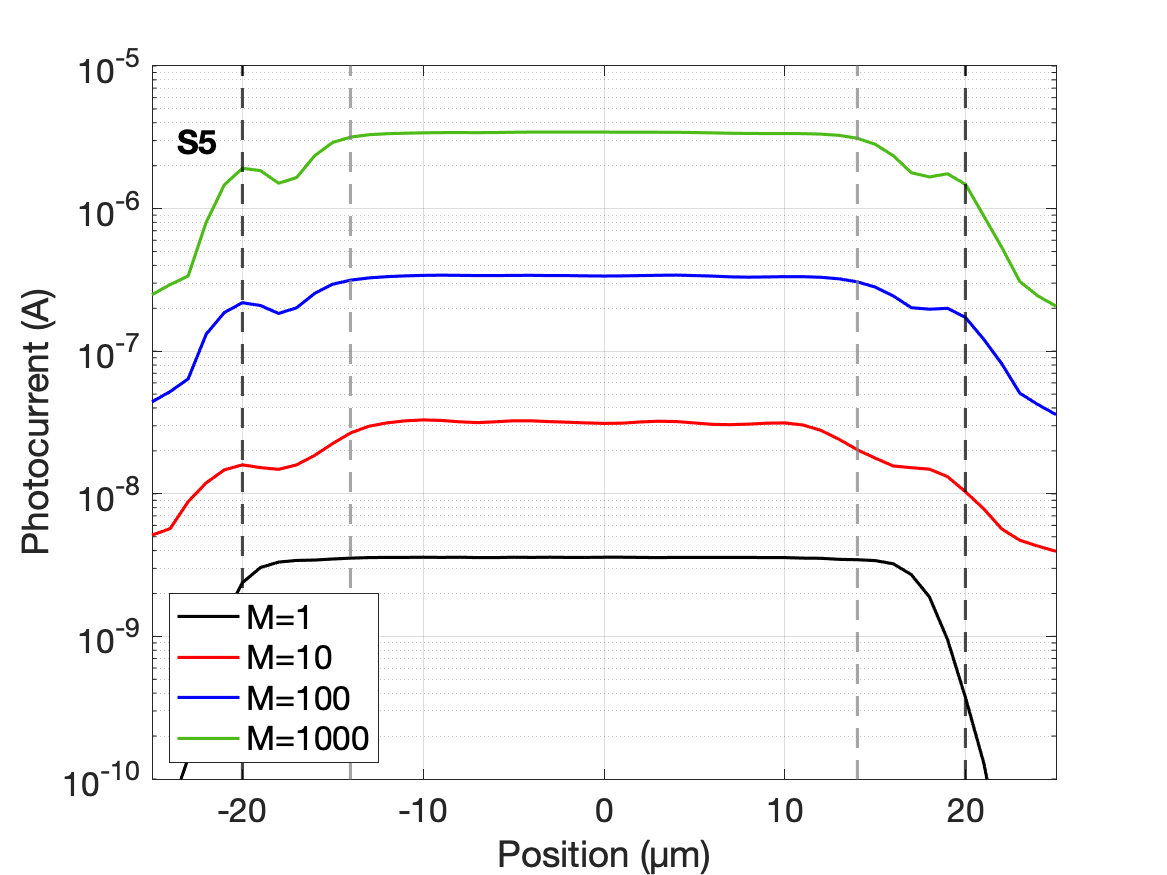}
\caption{}
\end{subfigure}
\begin{subfigure}{0.45\textwidth}
\includegraphics[width=\textwidth]{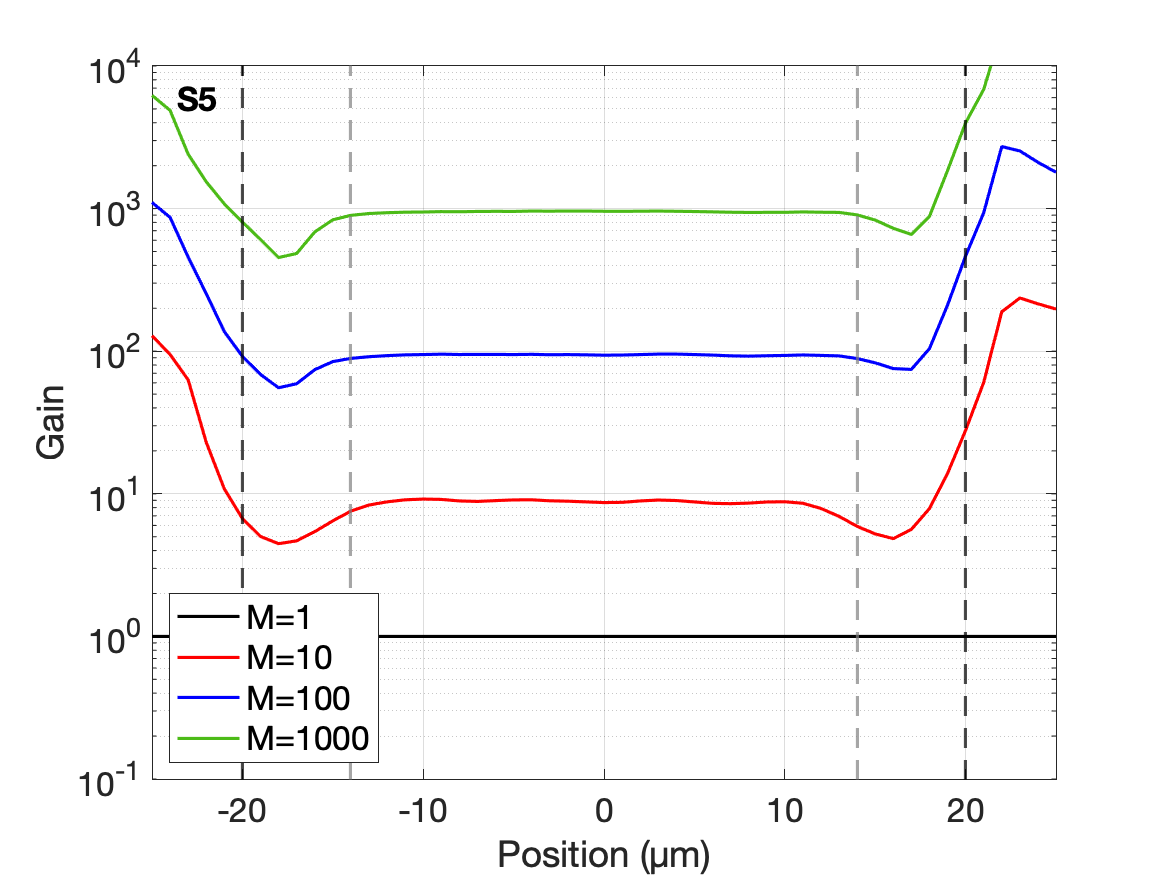}
\caption{}
\end{subfigure}
\caption{Measured photoresponse of CASSIA M2 (DP gain layer) with a n$^{+}$-electrode radius of 20$\,\rm\upmu$m and DP gain layer radius of 10$\,\rm\upmu$m (plots (a) and (b)) and of CASSIA S5 with an n$^{+}$-electrode radius of 20$\,\rm\upmu$m and DP gain layer radius of 14$\,\rm\upmu$m (plots (c) and (d)). The sensors are scanned by a laser beam with a FWHM of 2$\,\rm\upmu$m and 785\,nm wavelength, with M2 biased at reverse voltages of 30\,V (black curve), 39.4\,V (red curve), 41.1\,V (blue curve) and 45.1\,V (green curve) corresponding to gains of 1, 10, 100, and 1000, respectively. The respective voltages for S5 are 20\,V (black curve), 30.5\,V (red curve), 31.8\,V (blue curve) and 34.7\,V (green curve).}
\label{scan_m2_785}
\end{figure}

The photocurrent and gain distribution across CASSIA M2 and across CASSIA S5 with the same illumination setup, but with a laser wavelength of 785\,nm, are shown in Figure~\ref{scan_m2_785}. The bias voltages needed to achieve gains of 10, 100 and 1000 with the 785\,nm laser are slightly different than for the 532\,nm laser, and are equal to 39.4\,V, 41.1\,V and 45.1\,V, respectively for CASSIA M2 and 30.5\,V, 31.8\,V and 34.7\,V, respectively for CASSIA S5. Figures \ref{scan_m2_785}a and \ref{scan_m2_785}c give the dependence of photocurrent on radial beam position and \ref{scan_m2_785}b and \ref{scan_m2_785}d show the gain calculated as the measured photocurrent at the given voltages divided by the photocurrent without gain (black curve). CASSIA S5, shown in Figures \ref{scan_m2_785}c and \ref{scan_m2_785}d, has identical electrode/gain layer implantation but a gain layer radius of 14$\,\rm\upmu$m instead of 10$\,\rm\upmu$m (M2). For CASSIA M2 the photocurrent drops less sharply outside the DP layer from 10$\,\rm\upmu$m compared to 532\,nm illumination at a gain of 10. For the gains of 100 and 1000, the plateau in the active region with the highest gain is extended as compared to 532\,nm illumination, and photocurrent and gain drop more sharply. The absorption depth of light in silicon is around 10$\,\rm\upmu$m and 1.3$\,\rm\upmu$m at wavelengths of 785\,nm and 532\,nm, respectively. Hence, a lower portion of total light energy is absorbed between the n$^{+}$-electrode and DP gain layer, where the maximum electric field (Figure~\ref{tcad_field}a) and the corresponding high gain region (Figure~\ref{tcad_impact1}a) are located, in the case of 785\,nm illumination, and therefore a slightly higher voltage is needed to achieve the same gain as for 532\,nm illumination. A larger absorption depth for 785\,nm illumination implies a wider lateral light distribution in the silicon, meaning that more photons could be absorbed in the active region with high gain, resulting in a wider distribution of gain (Figure~\ref{scan_m2_785}b). Comparing results of CASSIA M2 with CASSIA S5 one can clearly observe the widening of the region with uniform gain when the gain layer radius is increased from 10$\,\rm\upmu$m to 14$\,\rm\upmu$m.

\begin{figure}[t!]
\centering
\begin{subfigure}{0.45\textwidth}
\includegraphics[width=\textwidth]{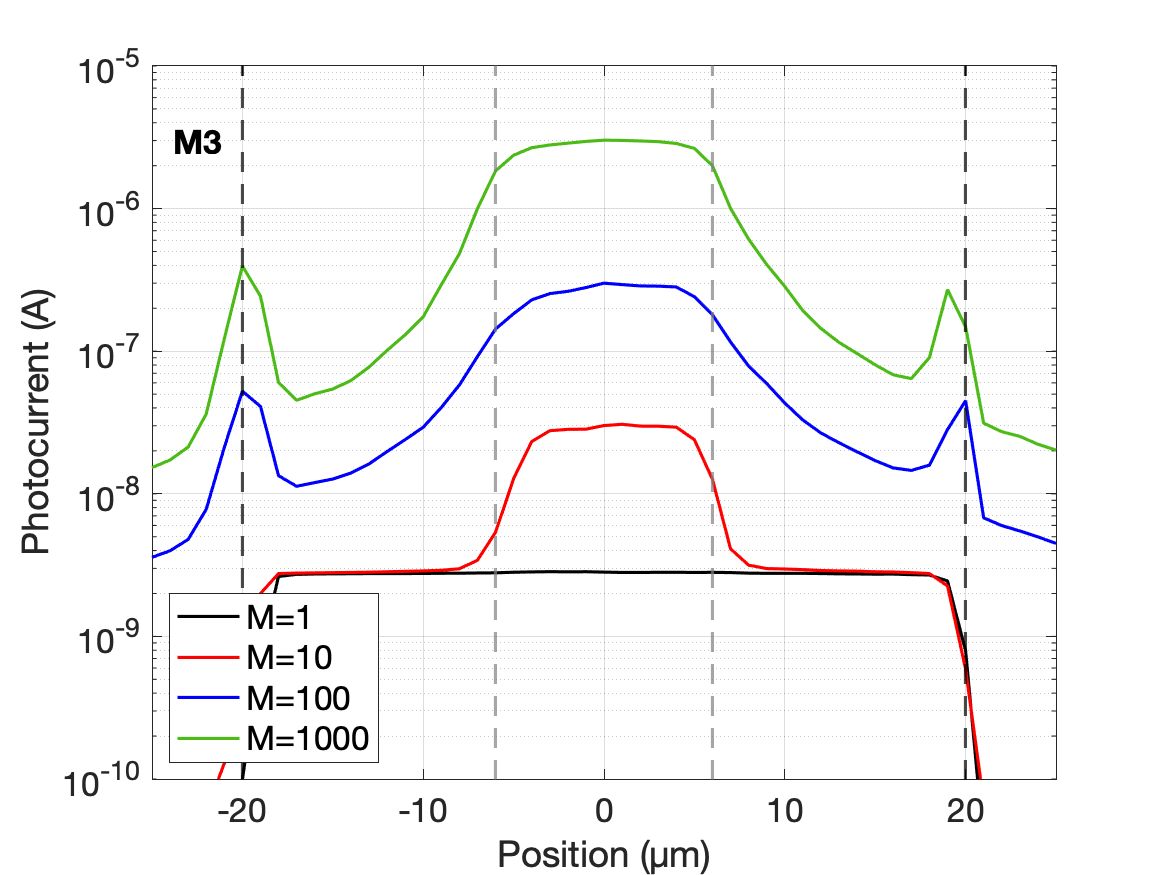}
\caption{}
\end{subfigure}
\begin{subfigure}{0.45\textwidth}
\includegraphics[width=\textwidth]{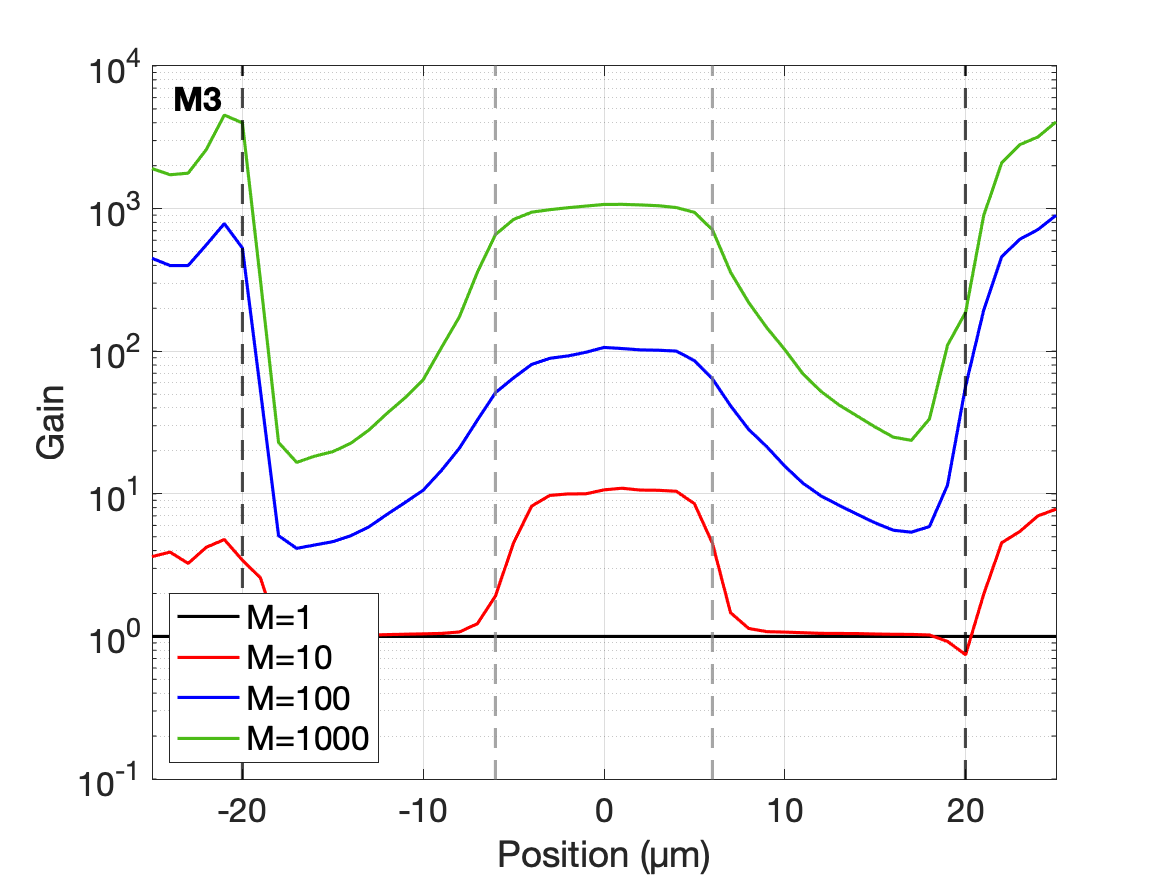}
\caption{}
\end{subfigure}
\caption{Measured photoresponse of CASSIA M3 (XDP gain layer) with an n$^{+}$-electrode radius of 20 um and XDP gain layer radius of 6$\,\rm\upmu$m scanned by a laser beam with a FWHM of 2$\,\rm\upmu$m and 532\,nm wavelength, biased at reverse voltages of 30\,V (black curve), 80.9\,V (red curve), 84.5\,V (blue curve) and 89.2\,V (green curve) corresponding to gains of 1, 10, 100, and 1000, respectively. (a) Dependence of photocurrent on radial beam position and (b) gain calculated as the measured photocurrent at the given voltages divided by the photocurrent without gain (black curve).}
\label{scan_m3_532}
\end{figure}

\begin{figure}[t!]
\centering
\begin{subfigure}{0.45\textwidth}
\includegraphics[width=\textwidth]{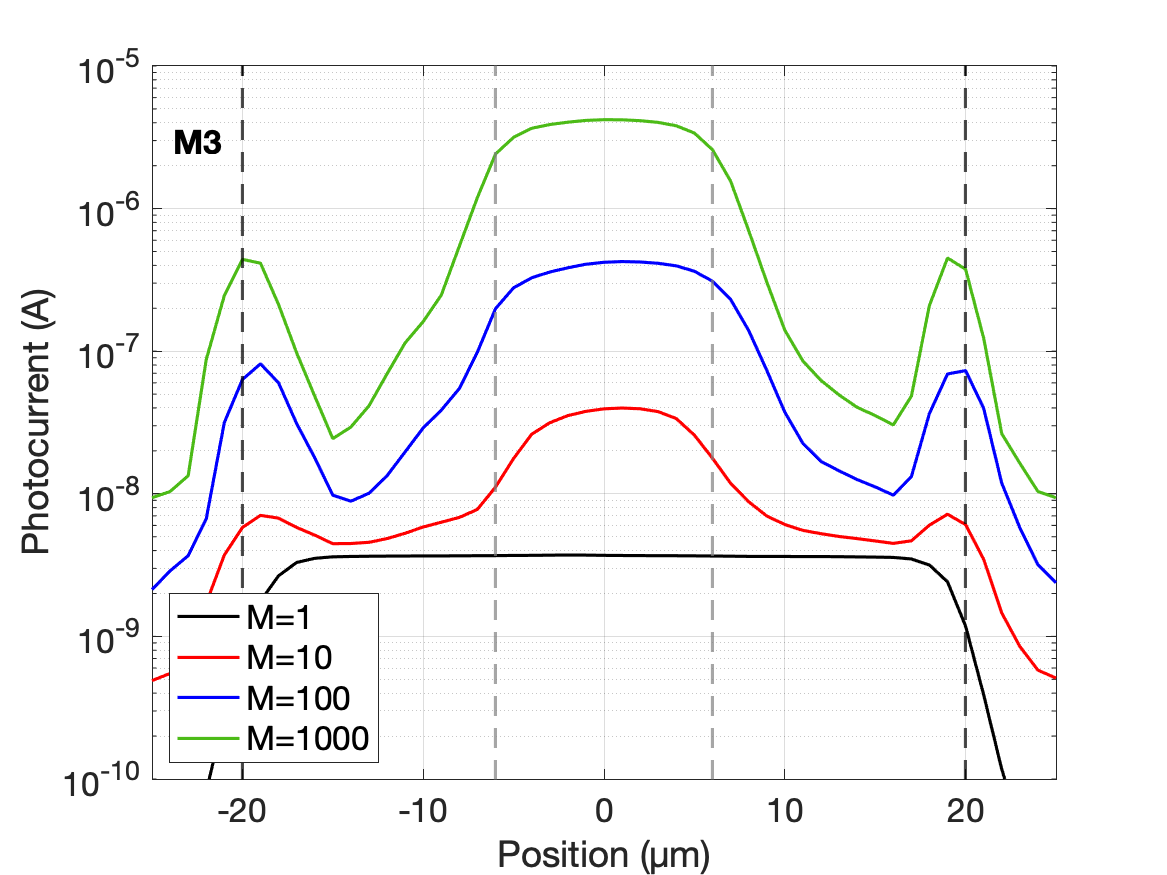}
\caption{}
\end{subfigure}
\begin{subfigure}{0.45\textwidth}
\includegraphics[width=\textwidth]{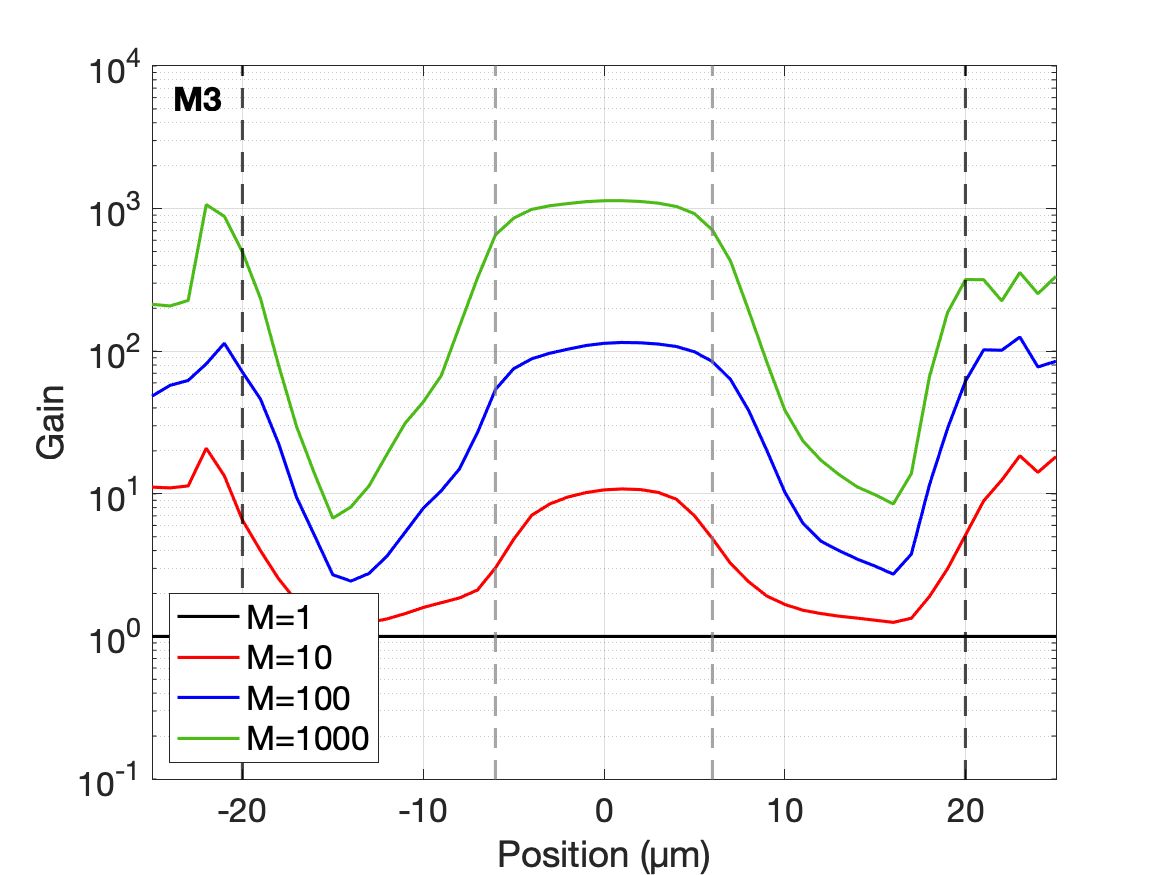}
\caption{}
\medskip
\end{subfigure}
\begin{subfigure}{0.45\textwidth}
\includegraphics[width=\textwidth]{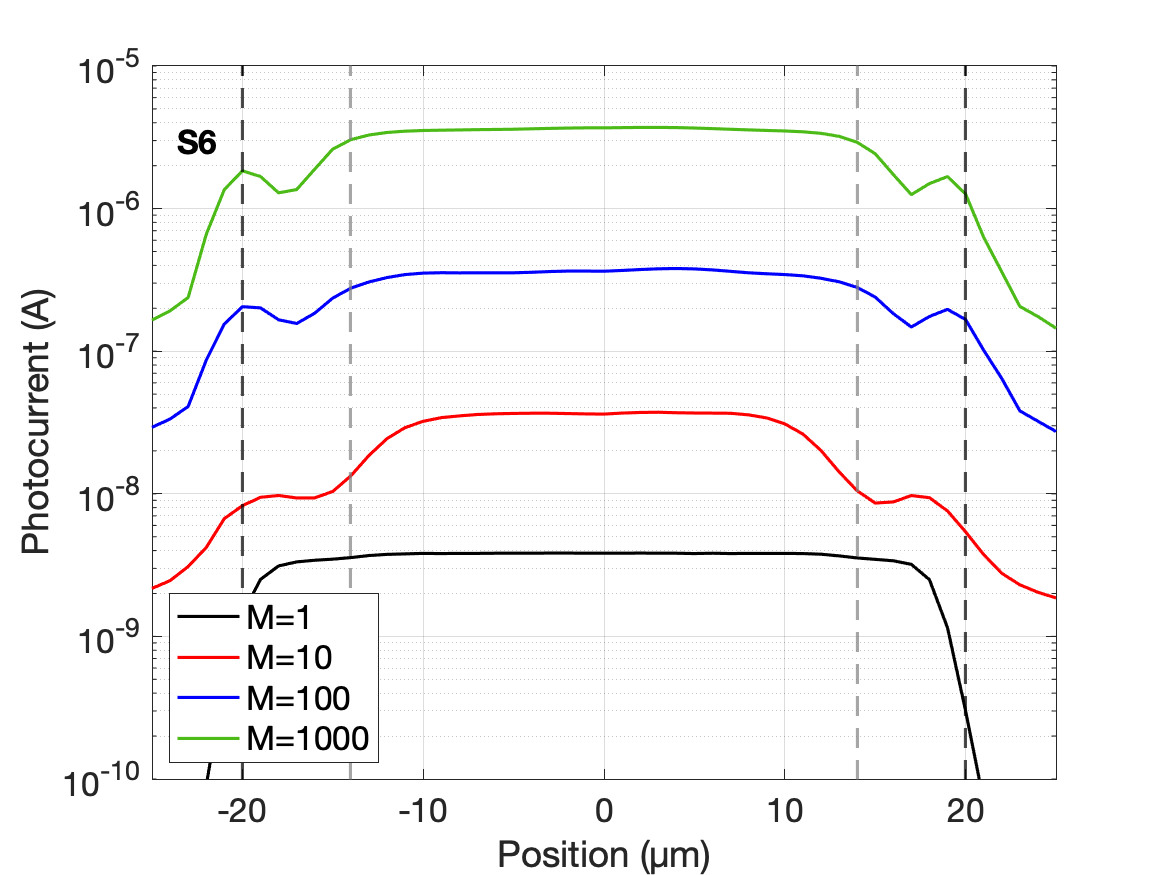}
\caption{}
\end{subfigure}
\begin{subfigure}{0.45\textwidth}
\includegraphics[width=\textwidth]{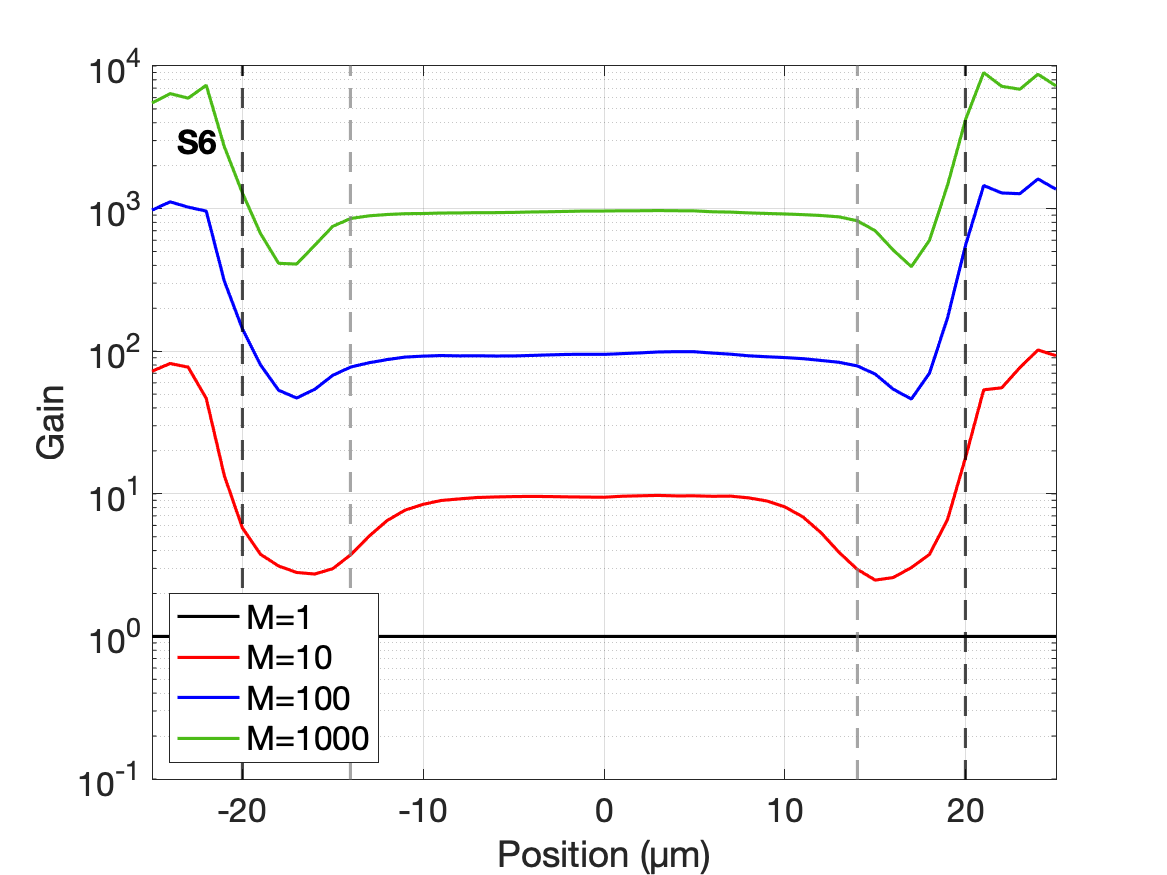}
\caption{}
\medskip
\end{subfigure}
\caption{Measured photoresponse of CASSIA M3 (XDP gain layer) with a n$^{+}$-electrode radius of 20$\,\rm\upmu$m and XDP gain layer radius of 6$\,\rm\upmu$m (plots (a) and (b)) and of CASSIA S6 with an n$^{+}$-electrode radius of 20$\,\rm\upmu$m and XDP gain layer radius of 14$\,\rm\upmu$m (plots (c) and (d)). The sensors are scanned by a laser beam with a FWHM of 2$\,\rm\upmu$m and 785\,nm wavelength, with M3 biased at reverse voltages of 30\,V (black curve), 82.4\,V (red curve), 85.2\,V (blue curve) and 90.7\,V (green curve) corresponding to gains of 1, 10, 100, and 1000, respectively. The respective voltages for S6 are 30\,V, 43.1\,V, 44.6\,V and 47.6\,V.}
\label{scan_m3_785}
\end{figure}

The measured dependence of photocurrent and gain on laser beam position in CASSIA M3, with an XDP layer radius of 6$\,\rm\upmu$m, for the wavelengths of 532\,nm and 785\,nm are shown in Figures~\ref{scan_m3_532}, ~\ref{scan_m3_785}a and ~\ref{scan_m3_785}b, respectively. A sharper drop of photocurrent and gain when moving away from the XDP gain layer region can be observed as compared to CASSIA M2 at both measured wavelengths. However, if the sensor is biased for a central gain of 100, it still has a gain above 10 at a radius of 11$\,\rm\upmu$m that is 5$\,\rm\upmu$m away from the XDP gain layer region. Similarly to CASSIA M2 when illuminated with the 785\,nm laser, CASSIA M3 requires a somewhat higher voltage to achieve the same gain as with 532\,nm illumination, since a smaller portion of light generates electrons between the n$^{+}$-electrode and (X)DP gain layer region. Furthermore, the photocurrent and gain exhibit a wider active-region plateau for 785\,nm illumination as compared to 532\,nm illumination, and then a sharper drop as the laser beam position moves towards the periphery. Figures ~\ref{scan_m3_785}c and ~\ref{scan_m3_785}d show the photocurrent and gain of CASSIA S6, which has the identical electrode/gain layer implantation as CASSIA M3 but a larger gain layer radius of 14$\,\rm\upmu$m. The increased gain layer radius results in a much wider region of uniform gain across the pixel.

The measurements presented in this section demonstrate that CASSIA sensors successfully achieve internal signal amplification through avalanche multiplication. The sensors exhibit distinct LGAD and SPAD operation modes, controllable by bias voltage, with uniform gain distribution across the active area. In the following section, we compare the performance of different pixel designs to identify optimal configurations for specific applications.

\section{Performance Comparison of CASSIA Designs}

To systematically evaluate the impact of different electrode and gain layer configurations on sensor performance, we conducted comparative measurements using pulsed laser illumination and dark count rate characterization. These measurements allow us to assess how design parameters such as gain layer depth, diameter and electrode implantation affect performance metrics including gain, breakdown voltages and noise.

\begin{figure}
    \centering
    \includegraphics[width=0.99\textwidth]{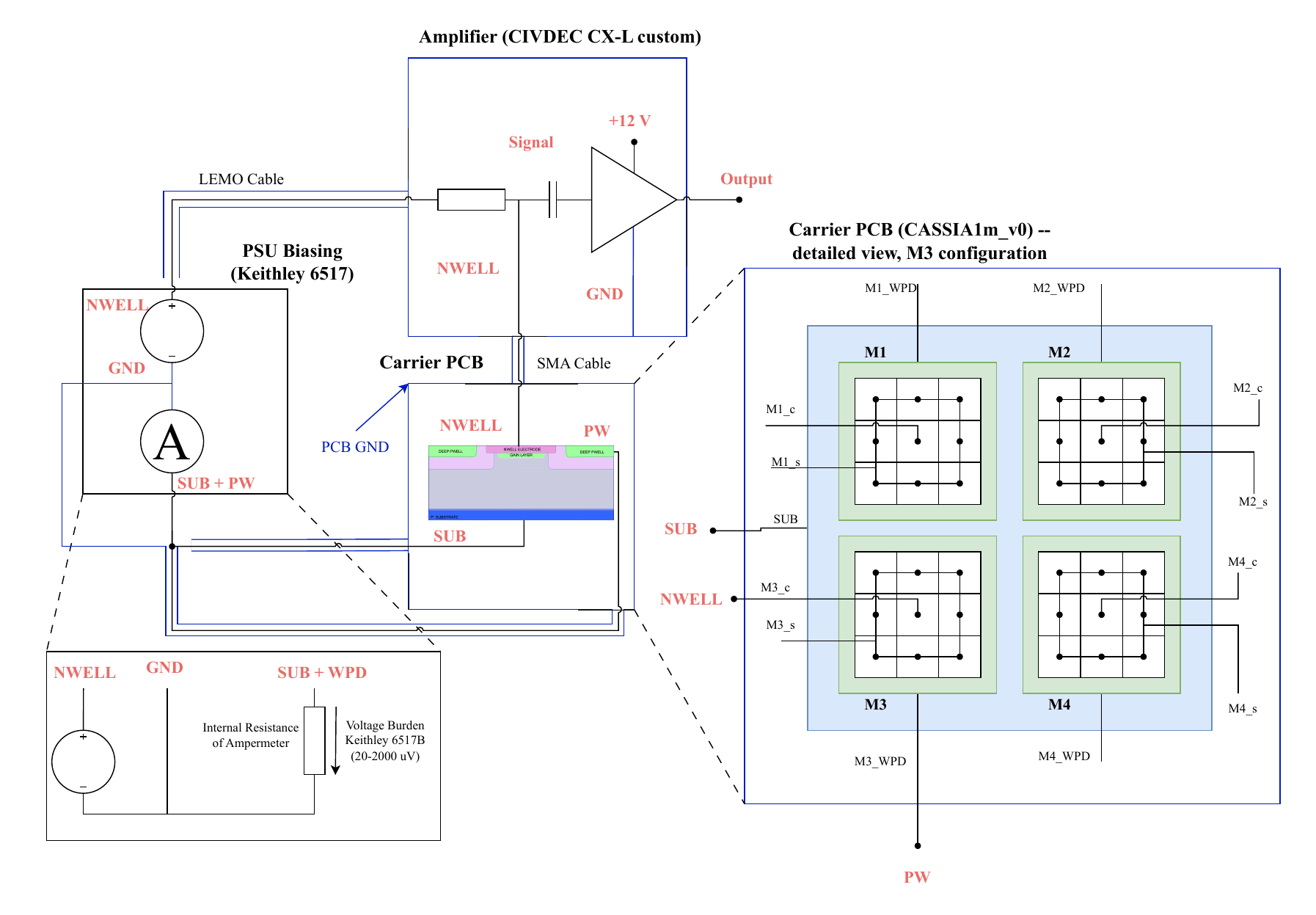}
     \caption{Setup used for CASSIA $I-V$, pulsed and DCR measurements. The CASSIA n$^{+}$-electrode is biased through the CX-L's internal bias-T. Substrate (SUB) and electronics deep p-well (PW) are connected to the PSU's ampermeter input (approximately 0V) for leakage current measurements ($I-V$s). The PCB's GND is connected to PSU GND to avoid measuring the PCB's leakage currents as the ampermeter is bypassed that way. The bias voltage provided from the PSU is therefore applied as voltage difference between electrode and substrate + electronics deep p-well.}
    \label{fig:CSchematic_Laser}
\end{figure}

For pulsed laser measurements the sensor is exposed to a triggered 1060nm PicoQuant\footnote{https://www.picoquant.com/products/category/picosecond-pulsed-sources} pulsed diode laser controlled by the PicoQuant PDL 800-B controller.  The pulse width is less than 100 ps. The laser is operated with a defocussed beam illuminating an area significantly larger than the matrix. The CASSIA pixel signal is routed via the PCB and SMA cable to an external CIVDEC Cx-L charge sensitive amplifier\footnote{https://cividec.at/index.php}. The setup to power and readout the CASSIA sensor during pulsed laser measurements is shown in Figure~\ref{fig:CSchematic_Laser}. The n$^{+}$-electrode is biased via an internal bias resistor of 300\,k$\Omega$. The Cx-L amplifier has a measured gain of 6.68\,mV/fC input charge, a rise time of 50\,ns to 70\,ns and FWHM pulse duration of $\approx$100ns.  It is designed to amplify positive and negative input signals with a linear input range of $\pm$150\,fC. The amplifier has an equivalent noise charge (ENC) of 338\,e$^{-}$ + 13e$^{-}$/pF. The r.m.s noise of the assembled CASSIA sensor on a PCB with signal routing on PCB and a SMA cable to amplifier is measured at operational conditions to be typically 2\,mV.

\subsection{Pulsed laser induced photo current on CASSIA sensors}  \label{sec:IV}

We carried out $I-V$ measurements with the laser off, and with the laser pulsed at different frequencies of 10\,kHz, 50\,kHz, and 100\,kHz. The vertical alignment of laser and sensor as well as the intensity of the laser has not been changed during the measurements to allow a relative comparison between the different matrices and single pixels. An X-Y position scan was carried out prior to each measurement to align the axis of the laser with the center of a single pixel, or the central pixel in a pixel matrix. Prior to the measurement on matrices with gain layers, the $I-V$ scan was carried our on CASSIA matrix M1 without gain layer. It showed pixel current of $\approx$1\,pA up to 160\,V bias on the electrode. CASSIA matrix M1 is identical for M2, M3 and M4 in all design parameters except the gain layer. With this measurement we verified that observed current breakdown is due to amplification in the gain layer and not due to e.g. breakdown to the electronics deep p-well or guard rings.

Break down values for all illuminated $I-V$  curves have been calculated using the method described in Section 4. Table~\ref{tab:breakdown} summarizes the mean of the break down value of the illuminated $I-V$  measurements for each of the matrices and single pixels.

\begin{figure}
    \centering
    \begin{tabular}{cc}
        \includegraphics[width=0.48\textwidth]{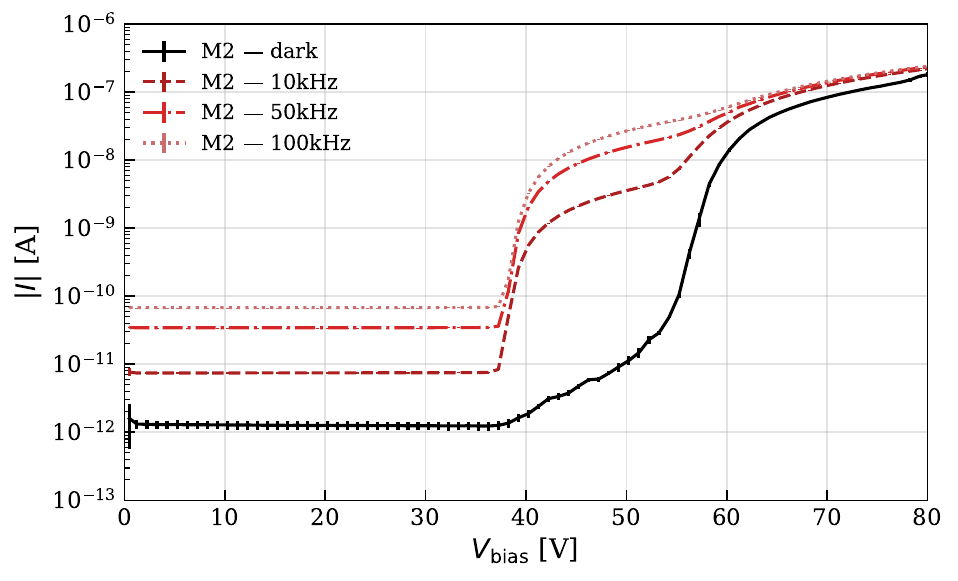}  &  \includegraphics[width=0.48\textwidth]{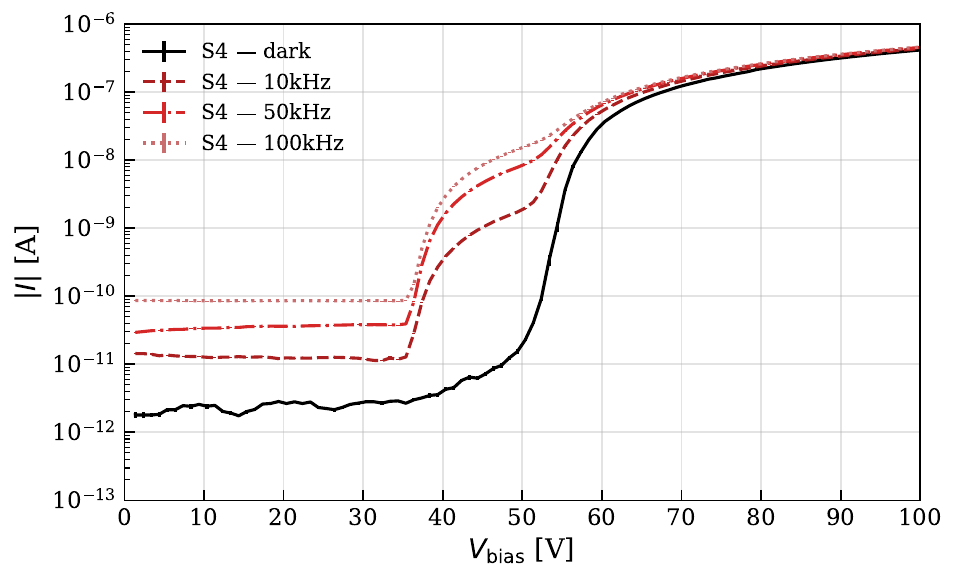}\\
       a) & b) \\
    
       \includegraphics[width=0.48\textwidth]{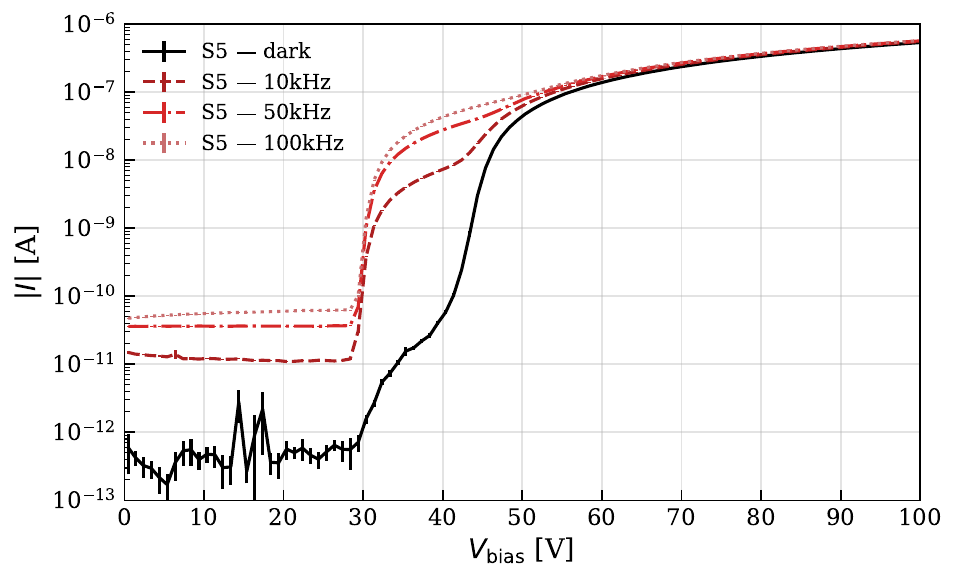}  &  \\
       c) &\\
    \end{tabular}
    \caption{Measured $I-V$  characteristics of CASSIA sensors with deep p-well (DP) gain layers under dark conditions and pulsed 1060\,nm laser illumination (10\,kHz, 50\,kHz and 100\,kHz): a) Matrix M2 (20$\,\rm\upmu$m DP gain layer diameter, standard spacing), b) Pixel S4 (20$\,\rm\upmu$m DP gain layer diameter, narrow n$^{+}$-electrode to electronics deep p-well spacing), c) Pixel S5 (28$\,\rm\upmu$m DP gain layer diameter, standard spacing). The horizontal axis shows bias voltage applied to the n$^{+}$-electrode.
    }
    \label{fig:NW-DPW-IV}
\end{figure}

CASSIA matrices significantly differ in their implant configurations and dimensions of gain layer diameter. For this reason we have grouped CASSIA matrices and single pixel measurements in groups of matching implant structures as described in Section~\ref{design} and Figure~\ref{fig:implants}: (a) M2/S4/S5, (b) M3/S6, (c) M4/S19 and (d) S13/S14. Each group shares identical implant configuration for n$^{+}$ electrode and gain layer but has different dimensions of the gain layer.  

Figure~\ref{fig:NW-DPW-IV} shows measured $I-V$  curves for CASSIA sensors M2 (Figure~\ref{fig:NW-DPW-IV}a), S4 (Figure~\ref{fig:NW-DPW-IV}b) and S5 (Figure~\ref{fig:NW-DPW-IV}b). The electrode is formed by the standard n$^{+}$-electrode and the gain layer by a DP well. M2 has a gain layer diameter of 20$\,\rm\upmu$m. S4 has a gain layer diameter of 20$\,\rm\upmu$m but a narrower spacing between electrode and electronics deep p-well. S5 has the identical design to M2 but with a gain layer diameter of 28$\,\rm\upmu$m. The figure shows as black curves the $I-V$  behaviour in dark measurements (laser off) and in red curves the $I-V$  measurements with the 1060nm laser pulsed at 10\,kHz, 50\,kHz and 100\,kHz respectively. Error bars in the plots denote the measurement's standard deviation. We observe that the current is in the pA range until low-gain mode amplification starts  at $\approx$38\,V on M2 in illuminated measurements. When the sensor is operated in LGAD mode, the observed current scales with increased frequency. Break down to strong amplification (SPAD mode) starts at about 55\,V for illuminated measurements. It should be noted that the current in SPAD mode is limited in this measurement to around 300\,nA not by the sensor but the external biasing resistance in the CIVIDEC amplifier circuit which supplies the bias voltage to the electrode. Figure~\ref{fig:NW-DPW-IV}c shows the same measurements on CASSIA S5 which has a gain layer diameter of  28$\,\rm\upmu$m. The increased gain layer size leads to a reduction of LGAD amplification voltage which is observed at 30\,V for S5. The larger gain layer area also leads to a significant increase in photo current amplification compared to M2 which is evident from $I-V$  curves at increasing pulse frequency.

\begin{figure}
    \centering
    \begin{tabular}{cc}
       \includegraphics[width=0.48\textwidth]{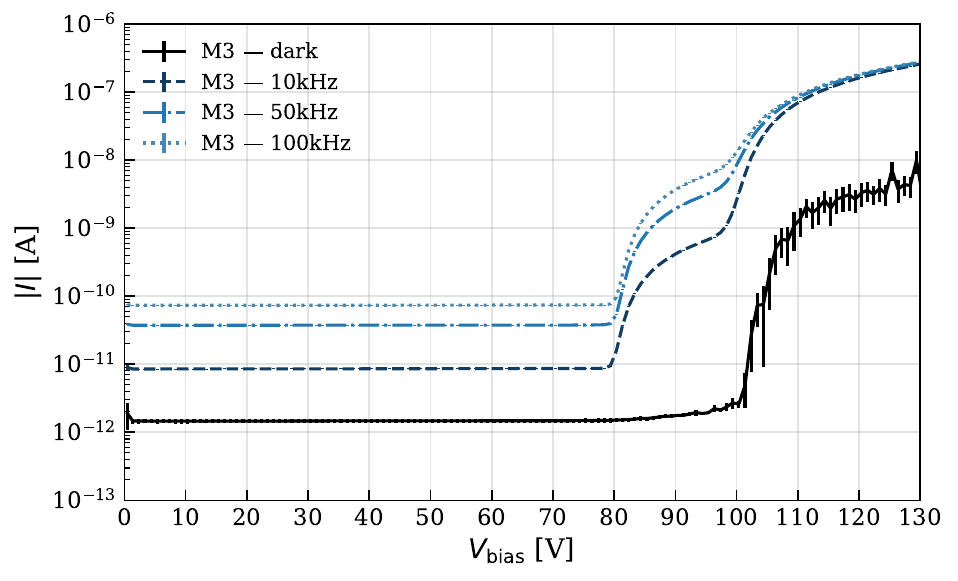}  &  \includegraphics[width=0.48\textwidth]{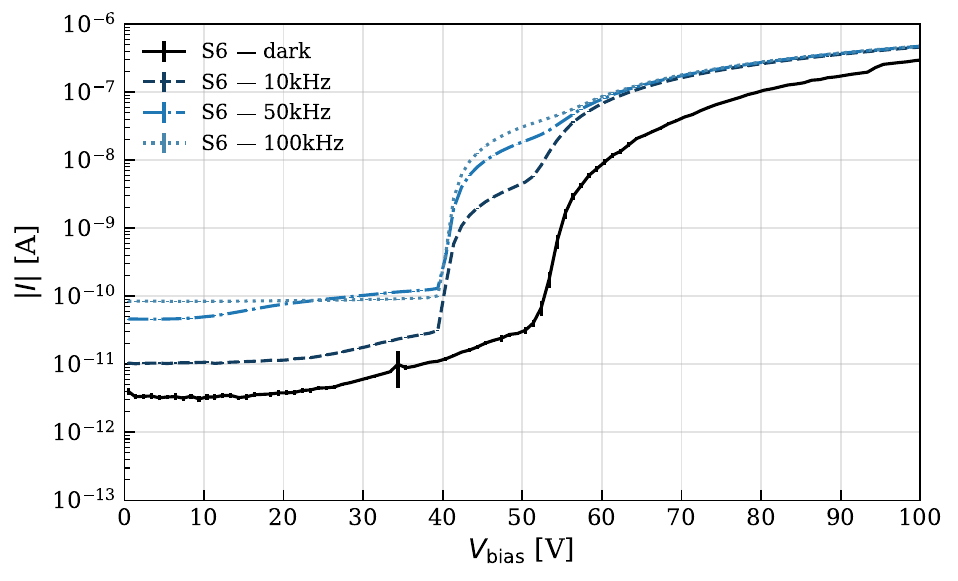}\\
       a) & b) \\
    
       \includegraphics[width=0.48\textwidth]{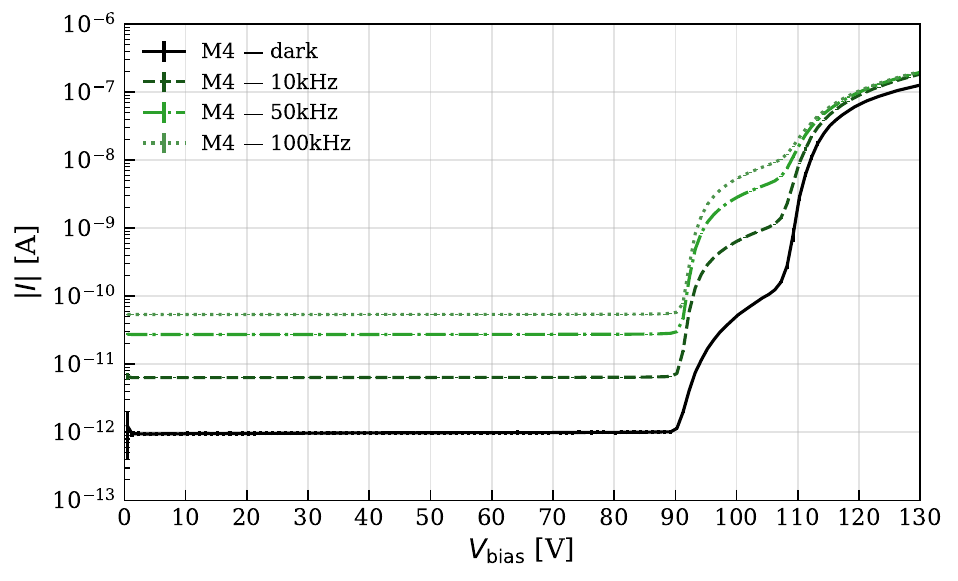}  &  \includegraphics[width=0.48\textwidth]{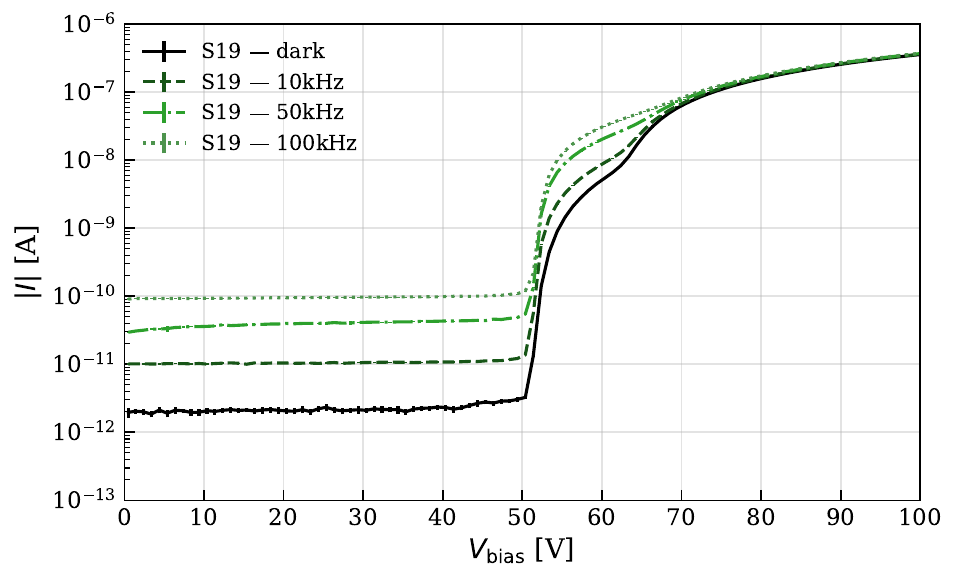}\\
       c) & d) \\
       
       \includegraphics[width=0.48\textwidth]{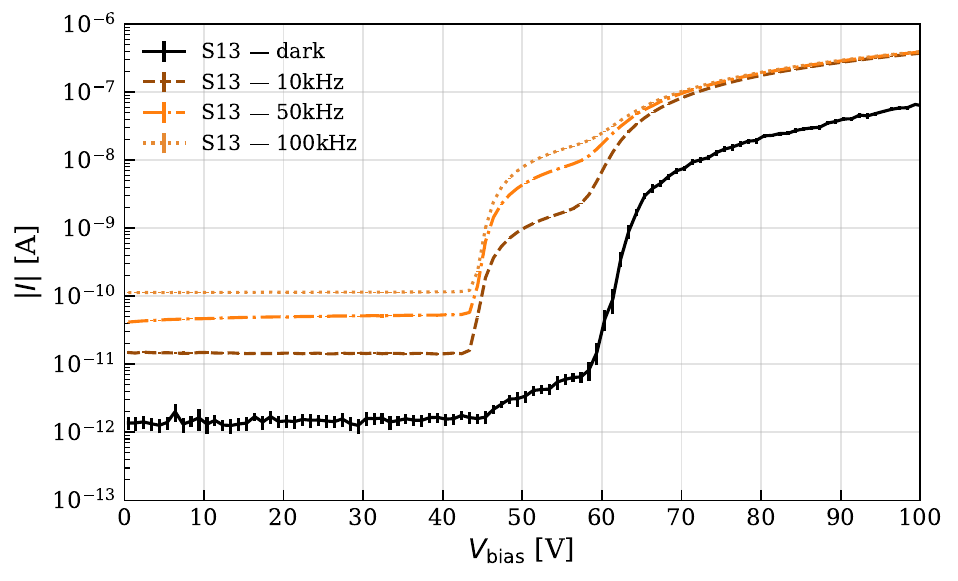}  &  \includegraphics[width=0.48\textwidth]{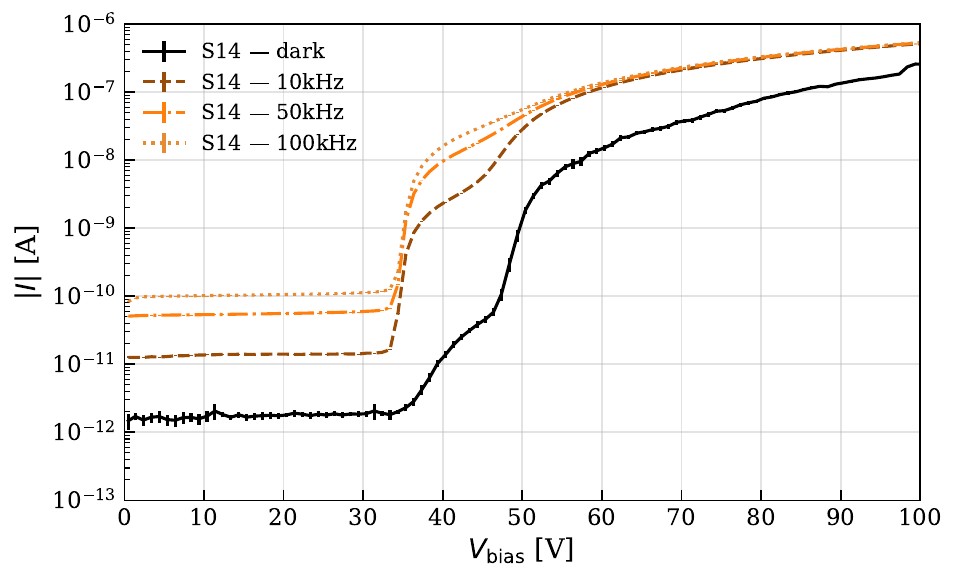} \\
       e) & f) 
    \end{tabular}
   
\caption{Measured $I-V$  characteristics of CASSIA sensors with extra deep p-well (XDP) gain layers under dark conditions and pulsed 1060\,nm laser illumination (10\,kHz, 50\,kHz, 100\,kHz): a,b) M3 (GL diameter 12\,$\upmu$m XDP) and S6 (GL diameter 28\,$\upmu$m XDP) with standard n$^{+}$-electrode, c,d) M4 (GL diameter 12\,$\upmu$m XDP) and S19 (GL diameter 28\,$\upmu$m XDP) with shallow n$^{+}$-electrode, e,f) S13 (GL diameter 20\,$\upmu$m XDP) and S14 (GL diameter 28\,$\upmu$m XDP) with deep n$^{+}$-electrode. The horizontal axis shows bias voltage applied to the n$^{+}$-electrode.}
   
    \label{fig:NW-XDPW-IV}
\end{figure}

Figure~\ref{fig:NW-XDPW-IV} shows measured $I-V$  curves for CASSIA sensors M3 (Figure~\ref{fig:NW-XDPW-IV}a) and S6 (Figure~\ref{fig:NW-XDPW-IV}b). The electrode is formed by a standard n$^{+}$-electrode and the gain layer by an XDP well. M3 has a gain layer diameter of 12$\,\rm\upmu$m  and S6 of 28$\,\rm\upmu$m. On CASSIA M3 low gain mode amplification occurs at 81\,V in illuminated measurements. At 100\,V the amplification shows a smooth transition to strong amplification in SPAD mode. On CASSIA S6, identical to M3 but with a large gain layer diameter, the measurements indicate that LGAD mode amplification already starts at 40\,V, less than half the voltage of M3. The difference in saturation current above breakdown for the dark and illuminated case is likely caused by the extremely low dark count rate in these devices. If dark counts occur very rarely, there is a possibility that an avalanche is not triggered even for the maximal measurement time attainable in our setup. Since the plots show the average value of multiple measurements performed during a given measurement time, this results in an apparent lower value of dark current above breakdown.

Figures ~\ref{fig:NW-XDPW-IV}c and ~\ref{fig:NW-XDPW-IV}d show measured $I-V$  curves for CASSIA sensors M4 and S19 respectively. The electrode is formed by a shallow n$^{+}$-electrode and the gain layer by an XDP well. 
M4 has a gain layer diameter of 12$\,\rm\upmu$m  and S19 of 28$\,\rm\upmu$m. LGAD mode amplification starts on M4 at 92\,V, it exhibits a smooth transition to stronger amplification (similar to M3 with the identical gain layer) at a voltage $\approx$110\,V. Breakdown occurs at 110\,V for pulsed measurements. The slight increase in breakdown from 100\,V on M3 to 110\,V at M4 can be attributed to the shallow electrode design in M4. Increasing the gain layer size to 28$\,\rm\upmu$m (sensor S19 in Figure~\ref{fig:NW-XDPW-IV}d) significantly lowers the onset of LGAD operation to 52\,V with a smooth transition to SPAD mode at 66\,V. Increasing the gain layer size also substantially increases the photo gain for the CASSIA design as can be seen in comparison to Figure~\ref{fig:NW-XDPW-IV}c.

While the comparison of CASSIA M3 to M4 analyses effects of a reduction in n$^{+}$-electrode depth, we designed CASSIA S13 and S14  sensors  with a deeper n$^{+}$-electrode implant than M3 while keeping the XDP gain layer. 
S13 has a gain layer diameter of 20$\,\rm\upmu$m  and S14 of 28$\,\rm\upmu$m. Figures ~\ref{fig:NW-XDPW-IV}e and ~\ref{fig:NW-XDPW-IV}f show measured $I-V$  curves for CASSIA sensors S13 and S14 respectively. With a gain layer diameter of 20$\,\rm\upmu$m the LGAD amplification starts at 45\,V, increasing the gain layer diameter to 28$\,\rm\upmu$m reduces the LGAD amplification start to 35\,V. Again the larger gain layer area leads to an increased photo gain. Like other CASSIA sensors using the XDP gain layer, we observe a smooth transition from LGAD mode to SPAD amplification.

\begin{table}
    \centering
    \begin{tabular}{|c|c|c|c|c|} \hline
       Sensor & Electrode & Gain layer (diameter) & $V_{\text{LGAD}}$ [V] & $V_{\text{BR}}$ [V] \\ \hline
        \hline
         M2 & NW & DPW (20$\upmu$m) & 38.8 & 58.0 \\ \hline
         M3 & NW  & XDPW (12$\upmu$m)  & 81.2 & 100.5\\ \hline
         M4 & shallow NW  &  XDPW (12$\upmu$m) & 92.3 & 109.6\\ \hline
         S4 & NW &  DPW (20$\upmu$m) & 36.8  & 54.0\\ \hline
         S5 & NW  &  DPW (28$\upmu$m) & 30.0 & 48.1 \\ \hline
         S6 & NW  &  XDPW (28$\upmu$m)  & 40.6 & 54.9\\ \hline
         S13 & deep NW & XDPW (20$\upmu$m) & 44.8 & 60.3\\ \hline
         S14 & deep NW & XDPW (28$\upmu$m) & 34.9 & 47.9\\ \hline
         S19 & shallow NW & XDPW (28$\upmu$m) & 51.9 & 66.4\\ \hline         
    \end{tabular}
    \caption{$V_{\text{LGAD}}$ and $V_{\text{BR}}$ as obtained in $I-V$  curves of pulses laser measurements for different CASSIA sensor designs.}
    \label{tab:breakdown}
\end{table}

\subsection{Amplitude distribution for LGAD and SPAD mode in different CASSIA designs}  

\begin{figure}
\centering
\includegraphics[width=1.0\textwidth]{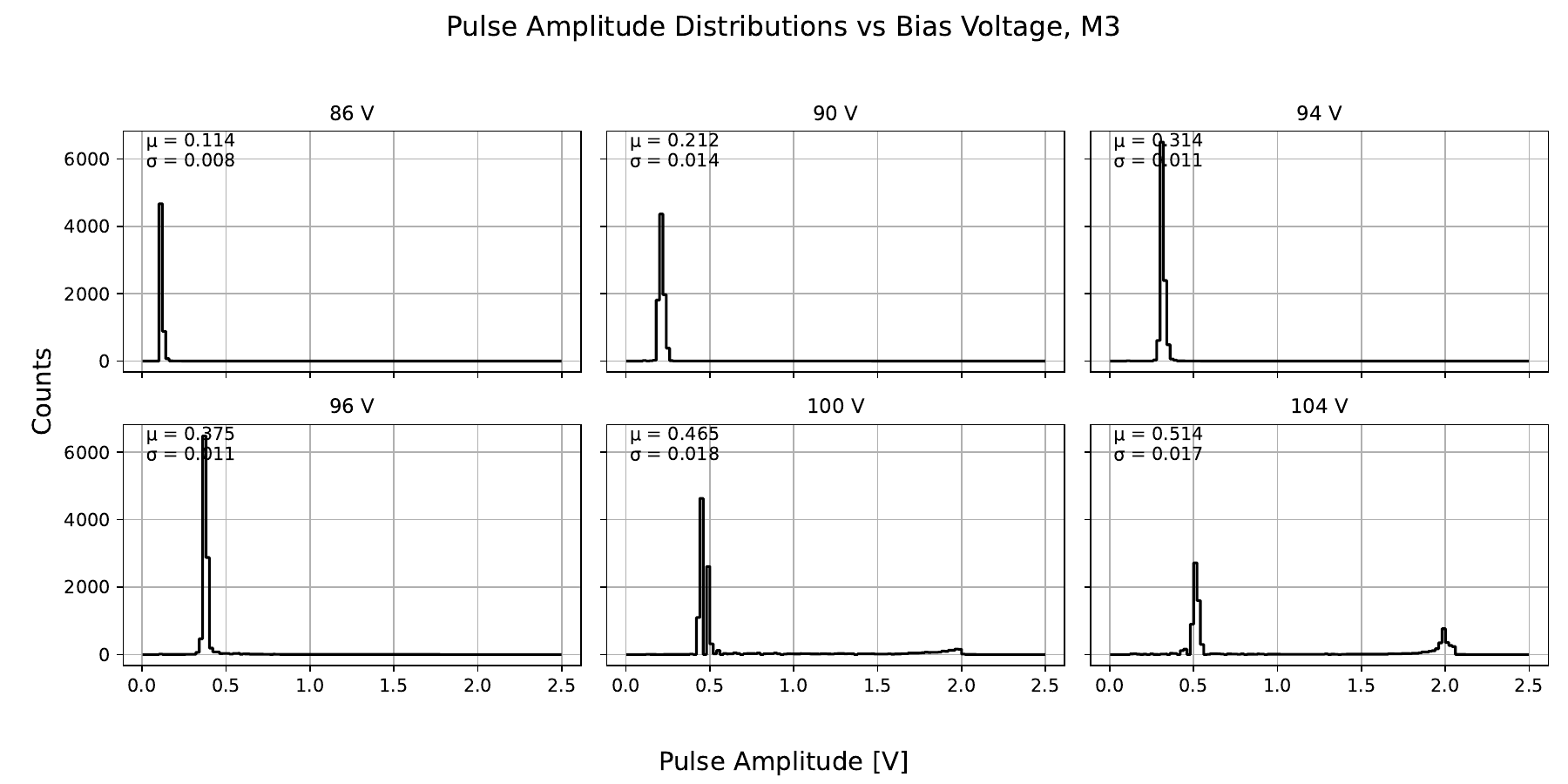}
\caption{Amplitude distribution of the central pixel of matrix M3 recorded at different cathode bias voltages with 1060nm triggered diode laser.
    }
    \label{fig:amp-dist}
\end{figure}

During the pulsed laser measurements, the amplifier output signal waveform of the central pad of the 3x3 matrix or the single pixel is recorded for each external trigger of the pulsed laser. The waveform is analysed offline where the start time of the pulse and peak amplitude are extracted from the waveform using a fixed threshold (typically $\approx$80mV). The time difference between pulse start-time and trigger $\Delta t$ is Gaussian distributed with an r.m.s spread of 0.6\,ns to 5\,ns depending on sensor bias voltage. This indicates a negligible contribution of random dark count pulses or after pulsing during the measurements. The r.m.s of $\Delta t$  is dominated by the external amplifier and cannot be used to extract the actual sensor time resolution. For each recorded waveform the peak amplitude is determined and histogrammed.
The amplitude distributions at different voltages for matrix M3 are shown in Figure~\ref{fig:amp-dist}. Taking matrix M3 of Figure~\ref{fig:amp-dist} as an example, we observe the onset of amplification at 86\,V, which correlates well with observation in $I-V$  measurements (Figure~\ref{fig:NW-XDPW-IV}a). The signal amplitude increases linearly with bias voltage at voltages above $V_{\text{LGAD}}$. The amplitude distribution matches a Gaussian distribution. At voltages below 100 V, the onset of avalanche multiplication is gradual and the internal amplification is limited like in an LGAD sensor. Above 100\,V, the avalanche multiplication becomes stronger resulting in a steeper current increase with reverse voltage and higher gain. The second peak in pulse response starts to occur, which corresponds to amplifier saturation and can be referred to as SPAD mode. The voltages around 100\,V correspond to the transition region from LGAD mode to SPAD mode, where we observe the parallel occurrence of both signals types.

We extract the mean amplitude of the LGAD-mode signal distribution with a Gaussian fit for each bias voltage from Figure~\ref{fig:amp-dist}. SPAD signals are too high for the amplifier dynamic range and result in a peak at 2V (corresponding to the maximum output voltage of the amplifier). The relative fraction of SPAD pulses to LGAD pulses rapidly increases with increasing voltage. Similar measurements and analysis has been carried out for M2, M4,S4, S5, S6, S13, S14 and S19. 

\begin{figure}
    \centering  
    \includegraphics[width=0.5\textwidth]{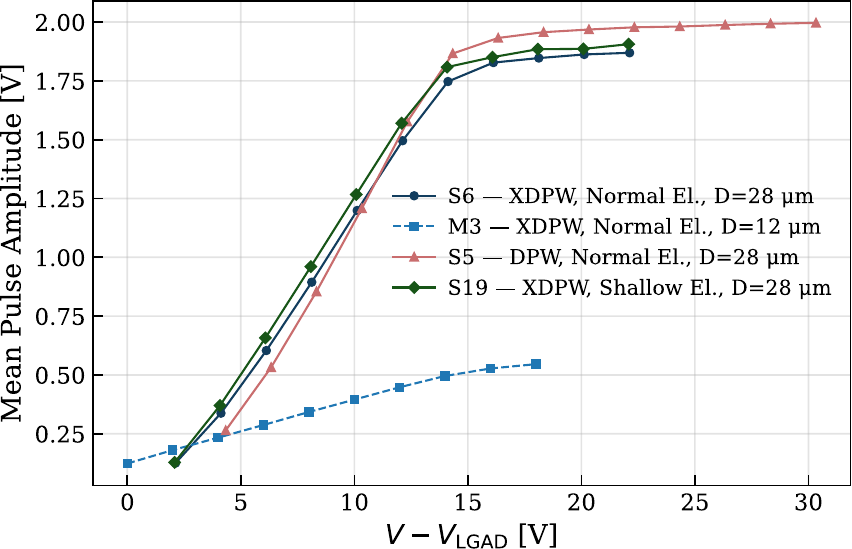}
    \caption{Mean LGAD-mode pulse amplitudes as a function of bias voltage minus $V_{\text{LGAD}}$ for different electrode/gain layer implant configurations.}
    \label{fig:Amp-vs-V}
\end{figure}

As we observed in dark and illuminated $I-V$  curve measurements, the onset of LGAD and SPAD operation voltage strongly depends on the gain layer diameter. Figure~\ref{fig:Amp-vs-V} shows the mean pulse amplitude as function of bias voltage minus $V_{\text{LGAD}}$. It shows a near linear increase of the LGAD signal amplitude as function of voltage. The horizontal axis gives the applied bias voltage minus the onset voltage for LGAD amplification $V_{\text{LGAD}}$. All structures with identical gain layer diameter, i.e. 28$\,\rm\upmu$m, show approximately identical increase of signal amplitude as function of voltage. The curve for M3 (dark blue, gain layer diameter 12$\,\rm\upmu$m) shows a lower gain than all other curves due to its smaller gain layer active area. The scaling corresponds to the ratio in gain layer area, hence photon flux per area. 

\subsection{Results of dark count rate measurements for different electrode/gain layer designs} 

As mentioned in the introduction, the dark count rate (DCR) is of critical importance for the operation of detectors in HEP experiments~\cite{spad_borna}. The actual specification on DCR will strongly depend on the experiment and operational conditions at the accelerator. Dark counts generate so-called "noise-hits" in the detector which can deteriorate detector performance at the level of pattern recognition and available data transmission bandwidth for particle hits. The DCR for CASSIA sensors is measured in the identical setup as pulsed laser measurements with CASSIA sensor connected to the CIVIDEC amplifier and the pixel electrode biased via the amplifier as shown in figure~\ref{fig:CSchematic_Laser}. In order to examine the CASSIA sensor at different temperatures we place the setup into a temperature controlled climate chamber. Care is taken to shield the assembly from environmental light which can affect the DCR measurement. For all sensors presented in the paper we carry out DCR measurements at 0$^{\circ}$C, 10$^{\circ}$C, 20$^{\circ}$C and 30$^{\circ}$C. Dark count rate is determined from the number of self-triggered waveforms at a threshold of $\approx$80\,mV which is significantly less than amplitudes at voltages when first LGAD pulses are detected. At the same time the threshold is significantly above the amplifier noise ($\approx$2\,mV r.m.s.) to exclude any noise pulses stemming from the amplifier.

\begin{figure}
    \centering
    \includegraphics[width=1.0\textwidth]{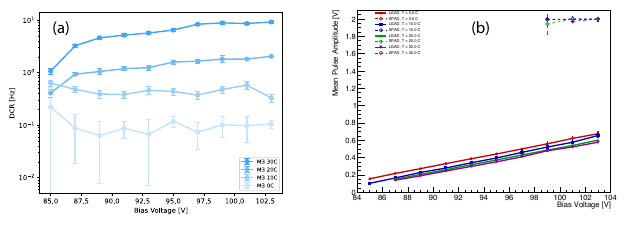}
    \caption{(a) Measured DCR for the central pixel of CASSIA matrix M3 operated at temperature of 0$^{\circ}$C, 10$^{\circ}$C, 20$^{\circ}$C and 30$^{\circ}$C. (b) Mean amplitude of M3 dark count LGAD pulses (solid line) and SPAD pulses (dashed lines) as function of voltage at different temperature. The horizontal axis gives the applied bias voltage to the n$^{+}$-electrode.
    }
    \label{fig:DCR-M3vsT}
\end{figure}

\begin{figure}
    \centering
    \includegraphics[width=1.0\textwidth]{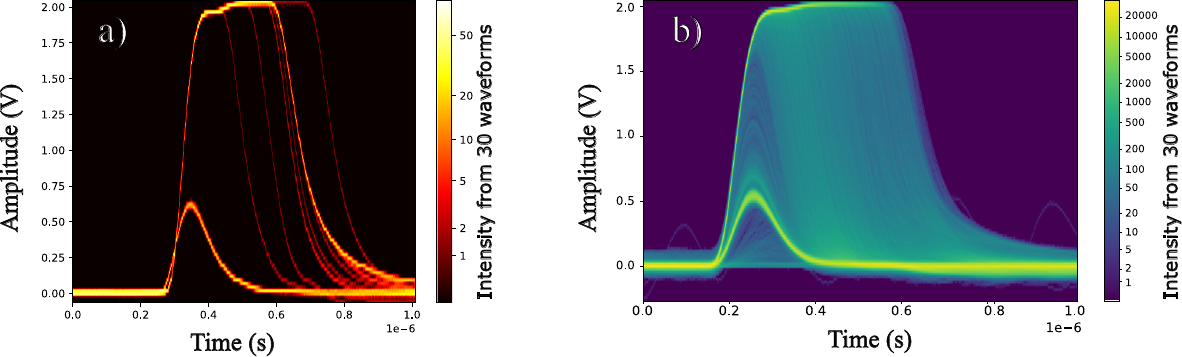}
    \caption{(a) Measured DCR waveforms for the central pixel of CASSIA matrix M3 operated at $V=$101\,V. (b) Measured waveforms of CASSIA matrix M3 dark for triggered 1060\,nm laser pulses with the matrix operated at $V=$104\,V.
    }
    \label{fig:DCR-laser-waveform}
\end{figure}

As example for the obtained results Figure~\ref{fig:DCR-M3vsT}a shows the DCR versus voltage at different temperature for CASSIA matrix M3. We observe an exponential increase of DCR with temperature which indicates noise stemming from thermally generated charge in the silicon sensor active volume~\cite{spad_mipro}. It should be noted that the absolute value is low at around 1Hz at room temperature in LGAD mode and early SPAD mode, which translates to $<$0.01 Hz/$\upmu$m$^{2}$ when normalized to the gain layer area. Increasing the voltage to 120V, i.e. exceeding $V_{\text{BR}}$ by 15\,V to operate in SPAD mode, the DCR increases to 0.1\,Hz/$\upmu$m$^{2}$. Figure~\ref{fig:DCR-M3vsT}b shows the mean amplitude of the dark count pulses as function of bias voltage at different temperatures. We observe in DCR measurements mean amplitudes which are comparable to the ones obtained in pulsed laser measurements. Figure~\ref{fig:DCR-laser-waveform} shows the comparison of waveforms measured on CASSIA matrix M3 in DCR measurements (a) and in response to externally triggered 1060\,nm laser pulses (b). Waveforms recorded in Figure~\ref{fig:DCR-laser-waveform}a are randomly distributed in time and recorded through a trigger on the waveform itself, waveforms recorded in Figure~\ref{fig:DCR-laser-waveform}b are triggered by the externally triggered laser pulse and are time correlated to the laser pulse. We observe nearly identical pulse amplitudes for randomly distributed dark counts and laser triggered pulses at voltages which are below SPAD operation voltages. In LGAD detectors, noise can be typically suppressed by applying a charge threshold above the signal generated by a single-electron hole pair. Our measurements suggest that thermally generated carriers generate output pulses of the same amplitude as light-generated ones, leading to false hit signals. The probability of such false pulses appears design specific and can reach extremely low values in structures with low DCR, such as the CASSIA sensor M3 shown in Figure ~\ref{fig:DCR-M3vsT}a, which is acceptable for typical tracking sensors. Coexistence of the pulses with different amplitudes might indicate two different mechanisms of amplification, as also observed in other LGAD sensors \cite{LASTOVICKAMEDIN2024169635}. Investigations to the possible origin of dark count pulses in LGAD mode are currently on-going.

\begin{figure}
    \centering
\includegraphics[width=0.9\textwidth]{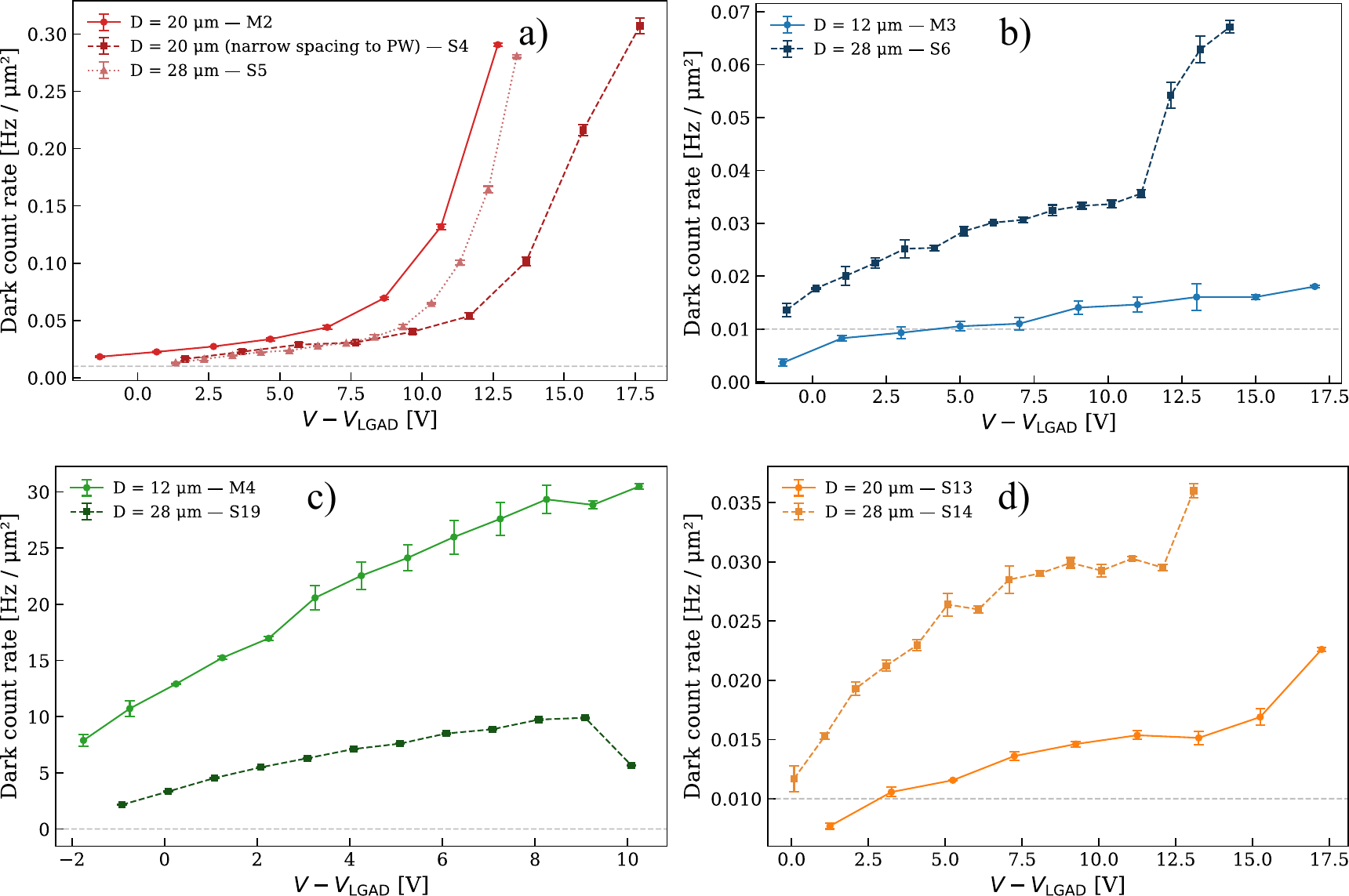}
    \caption{Normalized DCR at room temperature for CASSIA sensors with different electrode and gain layer designs. The horizontal axis shows the applied bias voltage to the n$^{+}$-electrode minus $V_{\text{LGAD}}$, the minimal bias voltage at which LGAD pulses appear. a) Standard n$^{+}$-electrode and DP gain layer (M2/S4/S5), b) standard n$^{+}$-electrode and XDP gain layer (M3/S6), c) deep n$^{+}$-electrode with XDP gain layer (M4/S19) and d) shallow n$^{+}$-electrode and XDP gain layer (S13/S14).}
    \label{fig:DCR-vs-GLdesign}
\end{figure}

For the comparison of DCR in different CASSIA designs we normalize the measured DCR to the gain layer area and present it as function of bias voltage minus the voltage when LGAD amplification starts which is denoted as $V_{\text{LGAD}}$ for this analysis. Figure~\ref{fig:DCR-vs-GLdesign} shows the normalized DCR as function of bias voltage minus $V_{\text{LGAD}}$. The measurements are taken at 20$^{\circ}$C. Figure~\ref{fig:DCR-vs-GLdesign}a shows DCR per gain layer area of CASSIA sensors using a standard n$^{+}$-electrode and a DP gain layer, Figure~\ref{fig:DCR-vs-GLdesign}b shows DCR per gain layer area of CASSIA sensors using a standard n$^{+}$-electrode and a XDP  gain layer, Figure~\ref{fig:DCR-vs-GLdesign}c shows DCR per gain layer area of CASSIA sensors using a shallow n$^{+}$-electrode and a XDP gain layer and Figure~\ref{fig:DCR-vs-GLdesign}d shows DCR per gain layer area of CASSIA sensors using a deep n$^{+}$-electrode and a XDP gain layer. 

The lowest DCR per gain layer at a level of $\approx$0.01 Hz/$\upmu$m$^{2}$ is obtained for CASSIA designs which combine a standard or deep n$^{+}$-electrode with a XDP gain layer. CASSIA sensors using shallow n$^{+}$-electrode yield the worst DCR results up to 30\,Hz/$\upmu$m$^{2}$. This is consistent with previous observations that moving the multiplication junction deeper into the silicon is beneficial for the DCR, since it reduces the contribution of defects inherently present at the silicon surface and at interfaces between the silicon and shallow trench isolation (STI) oxides~\cite{spad_borna}.

\section{Conclusions}

Design and fabrication of pixel sensors with internal signal amplification have been successfully integrated in a 180\,nm CMOS image sensor technology. The structures used for amplification, a combination of n$^{+}$-electrode with different p-well gain layer underneath, only uses layers already available in the process, without any increase in its complexity or fabrication parameter modifications. The first CASSIA sensor prototypes use deep p-well (DP) or extra deep p-well (XDP) layers under the n$^{+}$-electrode. Our results clearly indicate that the novel CASSIA pixel designs in the TowerSemiconductor 180nm imaging CMOS process can employ dedicated gain layers under the electrodes to achieve internal charge multiplication to amplify the primary ionization signal through impact ionization. The charge multiplication shows two distinct operation modes: a low-gain avalanche region first (LGAD) followed by a smooth transition to a high gain avalanche region where the detector operates like a SPAD. It is possible to operate the same pixel in either LGAD mode or SPAD mode simple by choice of electrode bias voltage. In LGAD mode low amplification of 10 to 100 is achieved over a range of typically 12V after amplification starts. The transition to SPAD operation is smooth and well controlled. In SPAD mode both gain layer type implementations (DP and XDP well) are suitable for achieving gains above 4000 with uniform distributions of the avalanche multiplication rate across the active area. Dark count rate measurements on sensors with different electrode/gain layer allowed to systematically study the influence on electrode and gain layer doping and depth on dark count rates. The measurements yield a low DCR of $\approx$0.01 Hz/$\upmu$m$^{2}$ at room temperature in the voltage range of interest for HEP applications. This work demonstrates the operation of CASSIA sensors with gain as an initial step of implementing large-scale monolithic sensors with internal gain for future applications in High Energy Physics and Nuclear Physics instrumentation.

\bibliographystyle{JHEP}
\bibliography{CASSIA.bib}

\end{document}